\documentclass[12pt]{article}
\usepackage{epsfig,amssymb,amsmath,psfrag}
\usepackage{cite}

\def\half{\textstyle{1 \over 2}}
\def\quarter{\textstyle{1 \over 4}}
\def\eps{\epsilon}
\def\eps{\epsilon}
\def\twome{ I^{\rm 2me}_{ij} }
\def\twomesix{ I^{\rm 2me,6D}_{ij} }
\def\One{{(1)}}
\def\Two{{(2)}}
\def\pel{{(\ell)}}
\def\xmp{ x^2_{i-1,i+1} }
\def\xij{ x^2_{ij} }
\def\wedge{W}
\def\wdiff{W_{i-1,i,i+1}}
\def\ump{  u_{i-1,i+1} }
\def\uij{  u_{ij} }
\def\suml{\sum_{\ell=1}^\infty}

\def\tf {  \tilde{f}  }
\def\tw {  \tilde{w}  }
\def\tC {  \tilde{C}  }
\def\tD {  \tilde{D}  }
\def\tF {  \tilde{F}  }
\def\tH {  \tilde{H}  }
\def\tI {  \tilde{I}  }
\def\tM {  {M}  }
\def\tR {  \tilde{R}  }
\def\tW {  {W}  }
\def\tPhi {  {\Phi}  }
\def\tG {  \tilde{G}  }
\def\tK {  \tilde{K}  }

\def\cG {  {\cal G}  }
\def\cN {  {\cal N}  }
\def\cO {  {\cal O}  }
\def\hx { \hat{x} }
\def\bi {\bar{\imath}}
\def\bj {\bar{\jmath}}
\def\al {\alpha}
\def\bet {\beta}
\def\dZ {{ d^4 Z_{\al\bet}\over i\pi^2 } }
\def\Li{ {\rm Li} } 

\def\viz{{viz.}}

\def\eqn#1{eq.~(\ref{#1})} \def\Eqn#1{Equation~(\ref{#1})}
\def\eqns#1#2{eqs.~(\ref{#1}) and~(\ref{#2})}

\def \bl  {\begin{align*}}
\def \el  {\end{align*}}
\def \be  {\begin{equation}}
\def \ee  {\end{equation}}
\def \ba  {\begin{eqnarray}}
\def \ea  {\end{eqnarray}}
\def \baa {\begin{eqnarray*}}
\def \eaa {\end{eqnarray*}}
\def \bb  {\begin {thebibliography} }
\def \eb  {\end{thebibliography}}
\def \lab #1 {\label{#1}}

\def \Tr {\mathop{\rm Tr}\nolimits}

\def \e  {\mathop{\rm e}\nolimits}

\newcommand \vev [1] {\langle{#1}\rangle}

\newcommand{\nn}{\nonumber}

\begin{document}

\begin{flushright}
HU-EP-11-41\\
DESY-11-148\\
BOW-PH-150\\
NSF-KITP-11-197\\
\end{flushright}
\vspace{3mm}

\begin{center}
{\Large\bf\sf  Form factors and scattering amplitudes in $\cN=4$ SYM\\
in dimensional and massive regularizations}

\vskip 5mm 
Johannes M. Henn\footnote{
{\tt henn@physik.hu-berlin.de
}}$^{,a,d}$,
Sven Moch\footnote{
{\tt sven-olaf.moch@desy.de
}}$^{,b,d}$, 
and 
Stephen G. Naculich\footnote{
{\tt naculich@bowdoin.edu
}}$^{,c,d}$ 
\end{center}

\begin{center}
$^{a}${\it
Institut f\"ur Physik\\
Humboldt-Universit\"at zu Berlin, 
Newtonstra\ss{}e15, D-12489 Berlin, Germany
}

\vspace{5mm}

$^{b}${\it 
Deutsches Elektronensynchrotron DESY,\\
Platanenallee 6, D-15738 Zeuthen, Germany
}

\vspace{5mm}

$^{c}${\it 
Department of Physics\\
Bowdoin College, Brunswick, ME 04011, USA
}

\vspace{5mm}

$^{d}${\it 
Kavli Institute for Theoretical Physics\\
University of California, Santa Barbara, CA  93106, USA
}

\end{center}

\vskip 2mm

\begin{abstract}
The IR-divergent scattering amplitudes of 
$\cN=4$ supersymmetric Yang-Mills theory 
can be regulated in a variety of ways, 
including dimensional regularization 
and massive (or Higgs) regularization.
The IR-finite part of an amplitude
in different regularizations generally differs 
by an additive constant at each loop order,
due to the ambiguity in 
separating finite and divergent contributions.
We give a prescription for defining an unambiguous,
regulator-independent finite part of the amplitude
by factoring off a product of IR-divergent ``wedge'' functions.
For the cases of dimensional regularization and 
the common-mass Higgs regulator, 
we define the wedge function in terms of a form factor,
and demonstrate  the regularization independence of the
$n$-point amplitude through two loops.
We also deduce the form of the wedge function for the
more general differential-mass Higgs regulator,
although we lack an explicit operator definition in this case.
Finally, using extended dual conformal symmetry,
we demonstrate the link between the differential-mass wedge function 
and the anomalous dual conformal Ward identity
for the finite part of the scattering amplitude.

\end{abstract}

\vfil\break

\section{Introduction and overview}
\setcounter{equation}{0}
\setcounter{footnote}{0}

Unraveling the pattern of soft and collinear divergences in scattering
amplitudes is a critical endeavor to advance our understanding of
gauge theories in general and to assist in concrete computations for
collider phenomenology, e.g., in massless gauge theories such as Quantum
Chromodynamics (QCD).  These studies have a long history (see e.g. the
early review~\cite{Collins:1989bt}) and have contributed to our knowledge
of the universal infrared (IR) structure of gauge theory amplitudes.
Based on the concepts of soft and collinear factorization, non-abelian
exponentiation, and the study of collinear limits, significant information
about scattering amplitudes is available to all orders in perturbation
theory.  For precision predictions at modern colliders, especially within
QCD and including higher order quantum corrections, these insights are
of great practical importance~\cite{Catani:1998bh,Sterman:2002qn}.

Quite generally, factorization implies the separation of scales in a
given scattering reaction, i.e. the process-dependent hard scale $q^2$
from those governing the soft and collinear limit, defined for instance
by the masses $m_i$ of the scattering particles with $q^2 \gg m_i^2$
or by the regions of soft momenta.  Note that the soft and collinear
singularities of massless particles (gauge bosons) strictly require
the definition of a regulator, which is conveniently performed in
$D=4-2\eps$ dimensions.  As an immediate consequence of factorization,
evolution equations emerge, which depend on the kinematics of the
specific process and on the chosen regulator.  Their solution leads to
non-abelian exponentiation, a result which also arises from an effective
field theory formulation based on the ultra-violet (UV) renormalization
properties of effective operators and their anomalous dimensions; see
e.g.~ref.~\cite{Collins:1989gx}.  Moreover, for scattering amplitudes in
gauge theories, the underlying factorization imposes strong constraints
on the anomalous dimensions and the all-order structure of the IR
singularities~\cite{Gardi:2009qi,Dixon:2009ur,Becher:2009cu,Becher:2009qa}.

In the present paper we will specialize our investigations in a number of
ways.  First, we choose to work in the $\cN=4$ supersymmetric Yang-Mills
(SYM) theory, which is the simplest non-abelian gauge theory in four
dimensions due to the vanishing of the four-dimensional $\beta$ function.
In our study, we are concerned with form factors and scattering amplitudes
in this theory, which allows us to study their IR singularities without
interference from UV divergences.  Second, we will be working in the
so-called planar limit and, for scattering amplitudes $A_n$ of $n$
external particles, we will assume color ordering.  Our main focus is on
the study of different kinematical regimes, i.e. scattering amplitudes
of massless and massive particles and associated form factors, 
using different regulator schemes.

The general property of factorization prompts us to ask whether one can
delineate a well-defined finite part of $A_n$ independent of the chosen
IR regularization.  While reasoning along these lines has already been
employed in the derivation of radiative corrections for heavy-quark
hadroproduction at two loops in QCD~\cite{Czakon:2007ej,Czakon:2007wk}
(see also refs.~~\cite{Ferroglia:2009ep,Ferroglia:2009ii}), this issue
is more generally related to the important question whether physical
observables in theories with massless particles are independent of
the regulator\footnote{ The well-known physical evolution kernels
are of course independent of the factorization scale $\mu^2$
by construction; see e.g. ref.~\cite{Soar:2009yh}.}; see e.g. the
discussion in ref.~~\cite{Contopanagos:1991yb}.  To that end, in this
paper we specifically compare dimensional and massive regularizations
schemes for $n$-particle scattering amplitudes $A_n$ in $\cN=4$
SYM theory, an ideal testing ground for these  questions due to the
simplicity of its loop expansion.\footnote{In the context of $\cN=4$
SYM theory, IR-safe inclusive differential cross-sections were studied
in ref.~\cite{Bork:2009ps}.} In-depth studies of the latter may lead
to new insights for gauge theories with massive particles which will
eventually also be of interest for applications in collider phenomenology.

Let us start thus, for simplicity, with maximally-helicity-violating (MHV)
scattering amplitudes.  Factorization implies that the color-ordered
amplitude $A_{n} = A_{n}^{\rm tree} M_{n}$ of $n$ particles can be
written as (see e.g.~ref.~\cite{Aybat:2006mz})
\begin{eqnarray}
\label{intro-factorization1}
M_{n} =  S_{n} \times J_{n} \times H_{n} \,.
\end{eqnarray}
Here $S_{n}$ and $J_{n}$ are ``soft'' and ``jet'' functions, respectively, and
$H_{n}$ is an IR-finite ``hard function''.
In general, $M_{n}$ and $H_{n}$ are vectors in a space of possible color structures, and $S_{n}$ is a matrix.
In the planar limit, $S_{n}$ is proportional to the identity matrix,
and one can combine $S_n$ and $J_n$ into a product of 
``wedge'' functions  $\wedge (s_{i-1,i})$ that depend only
on two adjacent particles $i-1$ and $i$ of the color-ordered amplitude \cite{Bern:2005iz}
\begin{eqnarray}
\label{intro-factorization2}
M_{n} = H_{n} \times \prod_{i=1}^n \wedge (s_{i-1,i}), 
\qquad\qquad 
s_{i-1,i} = (p_{i-1} + p_i)^2 \,.
\end{eqnarray}
As will be detailed below, the factorization
(\ref{intro-factorization2}) 
holds not only in dimensional regularization, 
but also in cases where masses 
are used to partially or fully regulate the IR divergences.
The wedge functions $\wedge (q^2)$ satisfy renormalization 
group equations which imply that they exponentiate. 
The factorization (\ref{intro-factorization2}) fixes
the hard function $H_n$            
only up to finite pieces, 
but with a suitable definition of the wedge function, we suggest that
\begin{eqnarray}
\label{log-hard-function}
\log H_{n} = \log M_{n} - \sum_{i=1}^n \log \wedge (s_{i-1,i})
\end{eqnarray}
can be used to define a regulator-independent finite 
part of the amplitude.

For regulators that leave the external particles massless,
such as dimensional regularization in $D = 4 - 2 \epsilon$ dimensions 
or the common-mass Higgs regulator described below, 
each wedge has half the IR divergences of a Sudakov form factor 
$\Phi(q^2)$ (see e.g. refs.~\cite{Korchemsky:1985xj,Dixon:2008gr}),
so it is natural to define
$\wedge (q^2) = \sqrt{\Phi(q^2)}$ \cite{Sterman:2002qn,Bern:2005iz}.
We show in this paper that, with this definition, $\log H_n$ is identical
for both of these regulators through two-loop order.

We also analyze a refined version of the Higgs regulator with 
differential masses, described below.
In this case, the external particles have distinct masses,
and so the wedge function cannot simply be defined in terms of
a Sudakov form factor.
Instead, we define the one-loop wedge function in terms of
a certain IR-divergent triangle diagram,
and then use extended dual conformal invariance to 
extend this to an all-loop expression for the 
sum of wedge functions.
With this choice for the IR-divergent wedge function, 
we establish that the IR-finite hard function $\log H_n$ 
takes precisely the same form for the differential-mass
Higgs regulator as for the common-mass regulator.
We lack, however, an explicit operator definition 
for the wedge function in this case.

For $\cN=4$ SYM theory, the regulator-independent hard function $\log H_{n}$
takes the simple form 
\begin{eqnarray}
\label{solution-Ward-identity}
\log H_n  =
{\gamma(a)\over 4}    H_n^\One  (\xij) + n D(a) + C(a) + R_n (\xij,a) 
\end{eqnarray}
due to the conjectured duality between 
the finite part of the MHV scattering amplitudes
and the (UV renormalized) expectation value of 
certain cusped Wilson loops
(see refs.~\cite{Alday:2008yw,Henn:2009bd} for reviews). 
In \eqn{solution-Ward-identity},
$\gamma(a)$ is the cusp anomalous dimension~\cite{Korchemskaya:1992je}, 
for which a prediction to all orders 
in the coupling constant $a=g^2 N/(8 \pi^2)$ exists~\cite{Beisert:2006ez},
and $D(a)$ and $C(a)$ are kinematic-independent functions.
The amplitude is expressed as a function of the dual or region momenta $x_{i}^{\mu}$,
which are defined by 
\be
p_{i}^{\mu} = x^{\mu}_{i} - x_{i+1}^{\mu}
\ee
with $x^\mu_{i+n} \equiv x^\mu_{i}$, and $x_{ij}^2 = (x_{i} -x_{j})^2$.
The first three terms on the r.h.s. of \eqn{solution-Ward-identity},
whose kinematical dependence is determined solely by the one-loop contribution $H_{n}^\One(\xij)$,
constitute the ABDK/BDS ansatz~\cite{Anastasiou:2003kj,Bern:2005iz}.
The {\it a priori} undetermined remainder function $R_{n}(\xij,a)$ contains 
the only non-trivial, i.e. loop-dependent, kinematical dependence.
\Eqn{solution-Ward-identity} follows from 
a conformal Ward identity for the dual Wilson 
loop \cite{Drummond:2007cf,Drummond:2007au}.
The first term on the r.h.s. of \eqn{solution-Ward-identity}  
provides a particular solution to this Ward identity.
The remainder function $R_{n}$ 
is the general homogeneous solution to the Ward identity,
and  depends only on dual conformal cross-ratios,
which take the form $x_{ij}^2 x_{mn}^2/(x_{im}^2x_{jn}^2)$.
Due to the absence of dual conformal cross-ratios for $n=4$ and $n=5$,
the remainder
functions $R_{4}$ and $R_{5}$ vanish and therefore the corresponding hard functions
$\log H_{4}$ and $\log H_{5}$ are completely determined by their one-loop value and
the kinematic-independent functions $\gamma(a), D(a)$, and $C(a)$.
For $n \ge 6$, dual conformal cross-ratios can be built, 
and the remainder function is known to be non-zero starting from
two loops and $n=6$ external particles \cite{Bern:2008ap,Drummond:2008aq}. Its higher-loop and higher-point
form is under intense investigation; see e.g. 
refs.~\cite{Drummond:2010cz,Gaiotto:2011dt,Bartels:2011nz,CaronHuot:2011ky,Dixon:2011pw}.

The planar MHV $n$-point amplitude for  $\cN=4$ SYM theory has
been studied using dimensional regularization and also 
using an alternative massive IR regulator\footnote
{ 
For earlier applications of a massive IR regulator, see
refs.~\cite{Alday:2007hr, Kawai:2007eg,Schabinger:2008ah,McGreevy:2008zy}.
}
\cite{Alday:2009zm,Henn:2010bk,Henn:2010ir,Drummond:2010mb,Craig:2011ws}.
The latter is motivated by the AdS/CFT correspondence and
consists of computing scattering amplitudes on the 
Coulomb branch of $\cN=4$ SYM theory,
i.e. giving a non-trivial vacuum expectation value to some of the scalars.
One can achieve a situation where the propagators on the perimeter of any
loop diagram are massive, thereby regulating the IR divergences.
The simplest case, the ``common mass Higgs regulator''
in which only one mass $m$ is introduced, 
corresponds to the breaking of the $U(N+M)$ gauge group to
$U(N) \times U(M)$, 
with fields in the adjoint representation of $U(M)$ remaining massless.

In the more general ``differential-mass Higgs regulator'',
one breaks the gauge group further to $U(N) \times U(1)^M $, 
thereby introducing various masses $m_{i}$, $i=1,\dots,M$. 
Fields in the adjoint of the broken $U(M)$,
which appear as external states in the scattering amplitudes,
now have nonzero masses $|m_{i}-m_{i+1}|\neq 0$. 
We use a decomposition of the one-loop MHV $n$-point amplitude
into a sum of IR-divergent triangle diagrams and 
IR-finite six-dimensional box integrals to
define the sum of one-loop wedge functions as\footnote{The subscripts
on the wedge function refer to its dependence on 
$m_{i-1}$, $m_i$, and $m_{i+1}$; see \eqn{wedge}.}
\ba
\label{sumwedgesoneloop}
\sum_{i=1}^n   \wdiff^\One  (\xmp)  \bigg|_{\rm one-loop}
=
\sum_{i=1}^n
\left[ - \frac{1}{4}
\log^2 \left( \frac{\xmp }{m_i^2} \right)
+ \frac{1}{8}  \log^2 \left( m_{i+1}^2 \over m_i^2 \right) 
\right]
\ea
in the uniform small mass limit
(i.e., $m_{i} = \alpha_{i} m$, with $\al_i$ fixed and $m \to 0$).
The one-loop hard function $H^\One_n (\xij)$ is then
expressed in terms of IR-finite quantities,
and thus is manifestly regulator-independent.

A key point is that the massive regulator is closely connected to dual conformal symmetry.
The Higgs masses can be interpreted within the AdS/CFT duality as the radial coordinates in a $T$-dual AdS${}_{5}$ space.
While the isometries of this space yield the usual dual conformal transformations for
zero masses, they define a different realization of this symmetry for finite masses,
dubbed ``extended dual conformal symmetry'' \cite{Alday:2009zm}.
Since no further regulator is needed in the massive setup,
the extended dual conformal symmetry is expected to be an
exact symmetry of the planar amplitudes.
Recently, it was shown that tree-level amplitudes on the Coulomb branch
of $\cN=4$ SYM 
and also all cuts of planar loop amplitudes do indeed have this 
extended symmetry~\cite{Dennen:2010dh,CaronHuot:2010rj}. 
Together with the expected cut-constructibility of $\cN=4$ SYM this then proves the 
extended dual conformal symmetry property conjectured in ref.~\cite{Alday:2009zm}.

What we wish to emphasize is that, while planar amplitudes
have extended dual conformal symmetry, the wedge functions
and regulator-independent hard functions separately do not.
This is not surprising, 
since extended dual conformal transformations act on the masses $m_i$ 
(as well as the dual variables $x_i$), 
whereas the hard function is,  by definition, independent
of the masses in the uniform small mass limit.
Nevertheless, extended dual conformal symmetry
can be used to determine the all-loop structure of the IR divergences 
of scattering amplitudes in the
case of the differential-mass Higgs regulator.
Assuming the validity of \eqn{bdshiggs}
for MHV scattering amplitudes, together with
\eqn{sumwedgesoneloop},
we obtain the following 
expression for the IR-divergent pieces in the differential-mass setup
\ba
\label{sumwedgesintro}
&&\sum_{i=1}^n   \log \wdiff (\xmp)  
\\
&&
\hskip10mm
=\sum_{i=1}^n
\left[ - \frac{\gamma(a)}{16} 
\log^2 \left( \frac{\xmp }{m_i^2} \right)
- \frac{\tilde{\cal G}_0(a)} {2}
\log\left( \frac{\xmp }{m_i^2} \right)
+ \frac{\gamma(a)}{32}  \log^2 \left( m_{i+1}^2 \over m_i^2 \right) 
+ \tw(a) 
\right]
\nn
\ea
valid for uniform small masses.

Having deduced the form of the IR divergences of 
the amplitude for the differential-mass regulator,
we turn the argument around and use \eqn{sumwedgesintro}
together with extended dual conformal symmetry to deduce
the anomalous dual conformal Ward identity,
from which the all-loop result (\ref{solution-Ward-identity}) 
follows.
Hence, a derivation of \eqn{sumwedgesintro} from first principles
would constitute a proof of \eqn{solution-Ward-identity}
without having to rely on the scattering amplitude/Wilson loop duality.
It would be very interesting to understand the origin of 
\eqn{sumwedgesintro} from a renormalization group approach.
A first step could be to find a suitable operator definition 
for the wedge function in the differential-mass regulator case. 
We leave these questions for future work.

This paper is organized as follows.
In section \ref{sect-bds}, we review the form of color-ordered MHV scattering amplitudes in planar $\cN=4$ SYM 
in dimensional regularization and in the massive regularization of ref.~\cite{Alday:2009zm}. 
In section \ref{sect-finite} we discuss factorization and exponentiation properties of scattering amplitudes and form factors.
We propose a definition,
involving Sudakov form factors, for a regulator-independent hard function 
that can be computed from the IR-divergent scattering amplitudes.
We review the result for the form factors up to two loops in dimensional regularization,
and compute the analogous quantities to two-loop order in the massive regularization.
We then show that the (logarithm of the) hard function defined earlier is the same in both cases.
In section \ref{sect-diff}, we discuss a more general differential-mass Higgs regularization,
and compute the IR-divergent terms of the one-loop amplitude in this regularization.
We then use extended dual conformal symmetry to derive the all-loop form of the IR-divergent 
terms, and discuss their relation to the dual conformal Ward identity.
Section \ref{sect-discussion} contains our conclusions, and
two appendices contain technical details used in the paper.

\section{Review of 
MHV amplitudes in $\cN=4$ SYM
}
\setcounter{equation}{0}
\label{sect-bds}

In this section, we briefly review the form of 
color-ordered MHV amplitudes in planar $\cN=4$ SYM theory
in different regularization schemes.

The all-loop-order $n$-point amplitude is given by the 
tree-level amplitude times a helicity-independent function
$M_n$, which we expand in the 't Hooft parameter
\be
\label{expansion}
M_n = 1 + \suml a^\ell M_n^\pel
,\qquad\qquad
a = {g^2 N \over 8 \pi^2} (4 \pi e^{-\gamma} )^\eps 
\ee
where $\eps = (4-D)/2$.
Loop-level amplitudes are UV-finite but suffer from IR
divergences which can be regulated using
either dimensional regularization in $D$ dimensions, 
or a Higgs regulator in four dimensions.
We discuss each of these in turn.

\subsection{Dimensional regularization of amplitudes}

In dimensional regularization, the $n$-point amplitude 
takes the form  \cite{Bern:2005iz}
\ba
\label{bds}
\log M_n  &=&
\suml   a^\ell
\left[ 
-\frac{\gamma^\pel}{ 8 (\ell \eps)^2} 
-\frac{\cG_0^\pel} {4 \ell \eps} 
\right]
\sum_{i=1}^n
\left( \mu^2 \over \xmp \right)^{\ell \eps}
\nonumber\\
&& \,+\, 
{\gamma(a)\over 4}    F_n^\One (\xij)
\,+\, n f(a) \,+\, C(a) \,+\, R_n (\xij, a)
\,+\, \cO(\eps) \,.
\ea
The momentum dependence of the amplitude 
is expressed in terms of dual variables $x_i$
defined via $x_i - x_{i+1} = p_i$,  
where $p_i$ are the momenta of the external states;
we also define $\xij  \equiv (x_i - x_j)^2$.
The terms on the first line of \eqn{bds} are IR-divergent
and are specified in terms of 
the cusp and collinear anomalous dimensions \cite{Korchemskaya:1992je}
\ba
\label{defgamma}
\gamma(a) 
&=& 
\sum_{\ell=1}^\infty   
\gamma^\pel a^\ell
= 
\sum_{\ell=1}^\infty   
4 f_0^\pel a^\ell
= 
4a - 4\zeta_2 a^2 
+ 22 \zeta_4 a^3 
+ \cO(a^4) \, ,
\\
\label{defG}
\cG_0(a)
&=& 
\sum_{\ell=1}^\infty    \cG_0^\pel a^\ell=
\sum_{\ell=1}^\infty    {2 f_1^\pel \over \ell} a^\ell=
-  \zeta_3 a^2 
+ \left(4 \zeta_5 + \frac{10}{3} \zeta_2 \zeta_3 \right)  a^3 
+ \cO(a^4) \,.
\ea
The terms on the second line of \eqn{bds} are IR-finite
and are determined by the finite part of the one-loop amplitude 
\be
\label{finiteoneloop}
F_n^\One  \equiv M_n^\One  
+ 
\frac{1}{2 \eps^2} 
\sum_{i=1}^n
\left( \mu^2 \over \xmp \right)^{\eps}
\ee
as well as the constants \cite{Bern:2005iz}
\ba
\label{deff}
f(a) 
&  =  &
\, -\, \suml \frac{ f_2^\pel}{2  \ell^2} a^\ell
=
 \frac{\pi^4}{720} a^2 
+\left( - \frac{c_1}{18} \zeta_6 - \frac{c_2}{18} \zeta_3^2 \right)  a^3 
+ \cO(a^4)\,,
\\
\label{defC}
C(a) 
&=& 
- \frac{\pi^4}{72} a^2 
+ \left[ 
\left( \frac{341}{216} + \frac{2 c_1}{ 9} \right)\zeta_6 
+ \left(- \frac{17}{9} + \frac{2 c_2}{ 9} \right)\zeta_3^2 \right]  a^3 
+ \cO(a^4)\,,
\ea
and a remainder function $R_n (\xij, a)$
potentially contributing beginning at two loops.
The original proposal by Bern, Dixon, and Smirnov \cite{Bern:2005iz}
conjectured that \eqn{bds} holds with $R_n (\xij, a)=0$. 
Explicit calculations bore this out for $n=4$ 
(through four loops) \cite{Bern:2006ew}
and $n=5$ (through two loops) \cite{Cachazo:2006tj},
but the two-loop calculation for $n=6$
\cite{Drummond:2007bm,Bern:2008ap,Drummond:2008aq}
revealed the necessity for a non-constant function 
$R_6(\xij,a)$. 

Explicit expressions for \eqn{finiteoneloop}
are given in ref.~\cite{Bern:2005iz}.
For $n=4$ and $n=5$, they are
\ba
\label{defF4}
F_4^\One &=&
\frac{1}{2} \log^2 \left( 
x^2_{13} \over x^2_{24} 
 \right) + \frac{2\pi^2}{3} \, ,
\\
\label{defF5}
F_5^\One &=&
 - \frac{1}{4} \sum_{i=1}^5
\log \left( \frac{ x^2_{i,i+2}}{x^2_{i+1,i+3} }\right)
\log \left(\frac{ x^2_{i-1,i+1}}{x^2_{i+2,i+4} }\right)
+ \frac{5\pi^2}{8} \,.
\ea

\subsection{Higgs regularization of amplitudes}

The four-, five-, and six-point functions have also been 
computed \cite{Alday:2009zm,Henn:2010bk,Henn:2010ir,Drummond:2010mb}
using the common-mass Higgs regulator
described in the introduction.
These amplitudes exhibit an exponentiation similar to \eqn{bds} 
which motivated the following analog for Higgs-regulated 
$n$-point amplitudes \cite{Alday:2009zm,Henn:2010ir}
\ba
\label{bdshiggs}
\log \tM_n &=& \sum_{i=1}^n
\left[ - \frac{\gamma(a)}{16} 
\log^2 \left( \frac{\xmp }{m^2} \right)
- \frac{\tilde{\cal G}_0(a)} {2}
\log\left( \frac{\xmp }{m^2} \right)
\right]
\nonumber\\
&& \,+\, \frac{\gamma(a)}{4} \, \tF_n^{(1)}  (\xij)
\,+\, n \tf (a)
\,+\, \tC (a)
\,+\, \tR_n  (\xij, a)
\,+\, \cO(m^2) \,. 
\ea
The terms on the first line of \eqn{bdshiggs} are IR-divergent.
The cusp anomalous dimension (\ref{defgamma}) is independent of
regularization scheme,
but the analog of the collinear anomalous dimension
is given by
\be
\label{defGt}
\tilde{\cG}_0(a)
=
- \zeta_3 a^2 
+ \left(\frac{9}{2} \zeta_5 - \zeta_2 \zeta_3 \right)  a^3 
+ \cO(a^4) \,. 
\ee
The terms on the second line of \eqn{bdshiggs} are IR-finite
and are determined by the finite part of the one-loop amplitude 
\be
\tF_n^\One \equiv M_n^\One  
+ \frac{1}{4}  \sum_{i=1}^n \log^2 \left( \frac{\xmp }{m^2} \right)
\ee
as well as the constants \cite{Henn:2010ir}
\ba
\label{defft}
\tf(a)  
&=&
  \frac{\pi^4}{180} a^2 + {\cal O}(a^3),
\\
\label{defCt}
\tC(a) 
&=&
 - \frac{\pi^4}{72} a^2 + {\cal O}(a^3) \,.
\ea
and a remainder function $\tR_n (\xij, a)$.
As in the case of dimensional regularization, 
the remainder function vanishes for four- and five-point amplitudes.
For $n=6$, it was shown \cite{Drummond:2010mb} 
that the two-loop remainder function $\tR_6^\Two(\xij)$
in the Higgs-regulated amplitude 
is precisely equal to its value $R_6^\Two(\xij)$
in dimensional regularization,
and this is expected to hold generally.

The one-loop amplitudes may be evaluated to show \cite{Henn:2010ir}
\ba
\label{defF4t}
\tF_4^\One &=&
\frac{1}{2} \log^2 \left(
x^2_{13} \over x^2_{24} 
\right) + \frac{\pi^2}{2}
= F_4^\One - \frac{\pi^2}{6} \,, 
\\
\label{defF5t}
\tF_5^\One &=&
 - \frac{1}{4} \sum_{i=1}^5
\log \left( \frac{ x^2_{i,i+2}}{x^2_{i+1,i+3} }\right)
\log \left(\frac{ x^2_{i-1,i+1}}{x^2_{i+2,i+4} }\right)
+ \frac{5\pi^2}{12}
= F_5^\One - \frac{5\pi^2}{24} \, ,
\ea
and more generally \cite{Drummond:2010mb}
\be
\label{FFt}
\tF_n^\One  = F_n^\One  - \frac{n \pi^2}{24}  \, .
\ee

\section{Defining a regulator-independent IR-finite amplitude} 
\setcounter{equation}{0}
\label{sect-finite}

Comparing the known expressions for Higgs-regulated
amplitudes (\ref{bdshiggs}) with those for dimensionally-regulated ones
(\ref{bds}),
one observes that the IR-finite parts of the amplitudes
are equal in both regularizations, up to constants.
In this section, we make the connection more precise 
by introducing a regulator-independent expression 
for the finite part of the amplitude.

In a planar theory, the factorization (see e.g.~ref.~\cite{Aybat:2006mz}) 
of color-ordered amplitudes takes the specific form \cite{Bern:2005iz}
\be
\label{eq:defMn}
M_n  =  \left[ \prod_{i=1}^n  \wedge (\xmp)  \right] H_n(\xij)
\ee
where $\wedge (\xmp)$ is an IR-divergent ``wedge function''
depending only on $(p_{i-1} + p_i)^2$
and resulting from the exchange of soft gluons in the wedge between 
the $(i-1)$th and $i$th external particles,
and $H_n (\xij)$ is an IR-finite hard function.
With a suitable definition for $\wedge(\xmp)$,
we can use 
\be
\label{finite}
\log H_n = \log M_n - \sum_{i=1}^n  \log \wedge (\xmp )
\ee
to define the IR-finite part of the amplitude.
The forms of both $M_n$ and $\wedge$ will depend on the 
specific regulator,
but we will find that $\log H_n$ is regulator-independent.

\subsection{Dimensional regularization of the form factor}

In dimensional regularization,
the wedge function can be defined as the square root of the gluon 
form factor \cite{Sterman:2002qn,Bern:2005iz}.
Form factors in $\cN=4$ SYM have been studied at strong 
coupling \cite{Alday:2007he,Maldacena:2010kp}, 
at one loop \cite{Brandhuber:2010ad} 
and at two loops \cite{vanNeerven:1985ja,Bork:2010wf},
while three-loop results can be inferred from the respective QCD computations 
\cite{Moch:2005id,Moch:2005tm,Baikov:2009bg,Lee:2010cga,Gehrmann:2010ue,Gehrmann:2010tu}
using the principle of maximal transcendentality;
see e.g. ref.~\cite{Kotikov:2004er}. 

In $\cN=4$ SYM we can equivalently use the form factor 
\be
\label{formfactor}
\Phi(q^2) = \langle  J, p_{i} |  \cO_{IJ} (q)   |  I, p_{i-1}  \rangle
\ee
for scalars $\phi_I$ coupling to the operator
\be
\cO_{IJ} =  \Tr \left[ \phi_I \phi_J -
 \frac{1}{6} \delta_{IJ} \sum_{K=1}^6 \phi_K \phi_K \right]
\ee
with $q^2 = (p_{i-1}+p_i)^2$.
The operator $\cO_{IJ}$ belongs to the stress-energy multiplet of $\cN=4$ SYM and
is not UV renormalized.
This form factor has been computed to two loops in
dimensional regularization \cite{vanNeerven:1985ja}
\begin{figure}[b]
\centerline{
{\epsfxsize3.5cm  \epsfbox{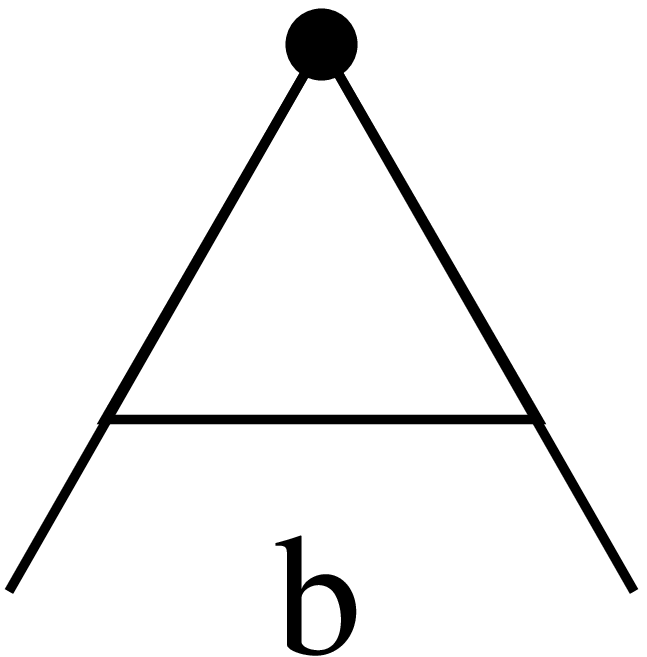}}
{\epsfxsize3.5cm  \epsfbox{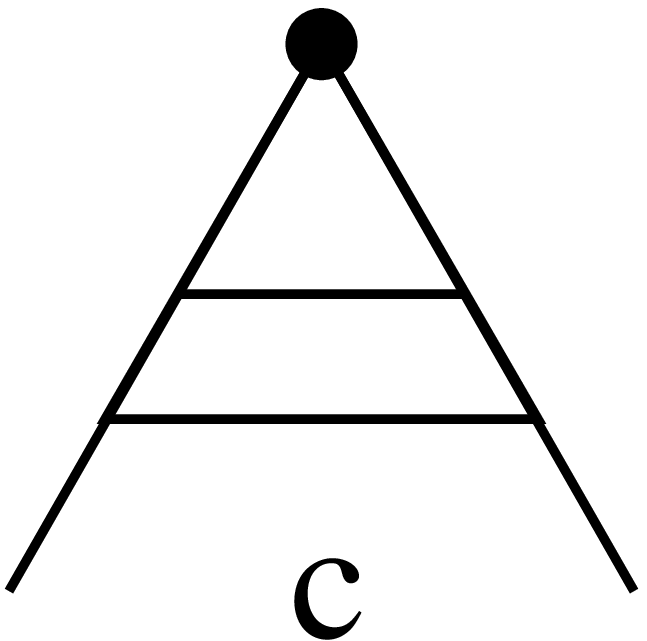}}
{\epsfxsize3.5cm  \epsfbox{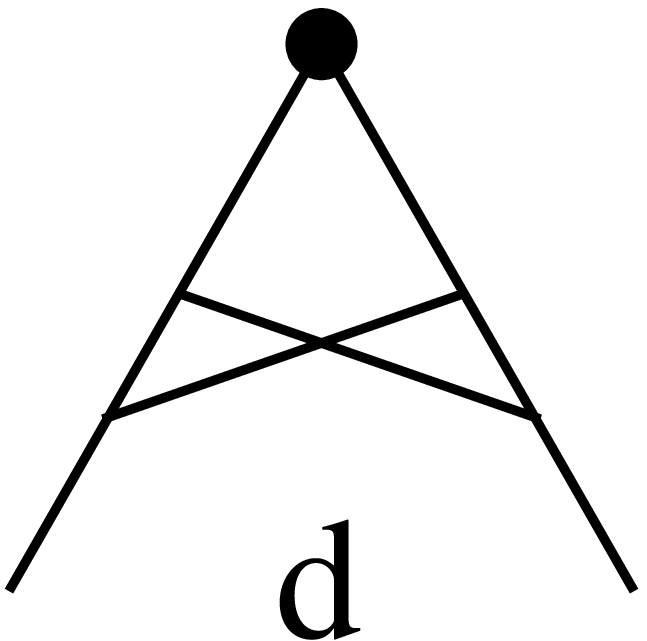}}
}
\caption{One- and two-loop form factor diagrams
in dimensional regularization.
All lines represent massless adjoint fields. 
The dot represents the insertion of $\cO_{IJ}$.}
\label{fig-massless}
\end{figure}
\be
\Phi(q^2) = 
1 + 
a q^2 \big[- I_b \big]
+ 
a^2 (q^2)^2 \big[ I_c + \quarter I_d \big] + \cdots
\ee
where $I_i$ represent the scalar integrals shown in fig.~\ref{fig-massless}.
(Despite its apparent non-planarity, integral $I_d$ is actually
leading order in the $1/N$ expansion as is clear 
from the corresponding double-line diagram in fig.~\ref{fig-doubleline}.)
The explicit expressions for these integrals 
given in appendix \ref{app-integrals}
reveal that the form factor exponentiates to two-loop order
\begin{equation}
  \label{eq:logPhi-2loop}
  \log \Phi(q^2) \,=\,
  a \left(\mu^2 \over q^2\right)^{\eps} 
  \left[ - {1 \over \eps^2} + {\pi^2 \over 12} + \cO(\eps) \right]
  + 
  a^2 \left(\mu^2 \over q^2\right)^{2\eps} 
  \left[  {\pi^2 \over 24 \eps^2} + {\zeta_3\over 4\eps} 
    + \cO(\eps) \right] 
  + \cO(a^3)\,.
\end{equation}

\Eqn{eq:logPhi-2loop} can be promoted to all orders in perturbation theory 
since the momentum dependence of $\Phi$ is governed by an evolution equation 
(see e.g. refs.~\cite{Korchemsky:1985xj,Dixon:2008gr,Korchemsky:1988hd,Moch:2005id}).
In $D=4-2\eps$ dimensions, the following factorization ansatz holds:
\begin{eqnarray}
  \label{eq:ffdeq}
  q^2 {\partial \over \partial q^2} \log \Phi(q^2)
  &=&
  {1 \over 2} \: K(a,\eps)
  + {1 \over 2} \: G(q^2/\mu^2,a,\eps) \, .
\end{eqnarray}
All dependence of $\Phi$ on the hard momentum $q^2$, which is taken to be Euclidean
($q^2>0$ in our mostly-plus metric convention) 
here and in the sequel, rests inside the function $G(q^2/\mu^2,a,\eps)$. 
The latter is finite in four dimensions and can be considered as a suitable continuation 
of the collinear anomalous dimension (\ref{defG}) to $D=4-2\eps$ dimensions.
$K(a,\eps)$ on the other hand serves as a pure counterterm.
Renormalization group invariance of $\Phi$ yields 
\begin{eqnarray}
  \label{eq:KGdeq}
  - \mu {d \over d \mu} K(a,\eps) \, = \,
  \mu {d \over d \mu} G(q^2/\mu^2,a,\eps) 
  \, = \,  
  \gamma(a) 
\end{eqnarray}
i.e., $G$ and $K$ are both governed by the cusp anomalous dimension 
$\gamma(a)$ of \eqn{defgamma}.
The solutions of \eqns{eq:ffdeq}{eq:KGdeq} can be conveniently given with the 
help of the $D$-dimensional continuation $\bar a$ of the 't Hooft parameter defined 
in \eqn{expansion}
\begin{eqnarray}
\label{eq:ad} 
{\bar a}(q^2,\eps) &=& 
a\, 
\left({\mu^2 \over q^2}\right)^{\eps}
\end{eqnarray}
which exhibits scale dependence on dimensional grounds and vanishes in the IR for $D=4-2\eps$ with $\eps < 0$.
Using \eqn{eq:ad} and exploiting the fact that $K$ has no explicit scale
dependence, which allows one to express it entirely through $\gamma(a)$,
one arrives at the following all-order expression for $\Phi$ 
\begin{eqnarray}
\label{eq:masslessPhi}
    \log \Phi(q^2)
    &=&
    {1 \over 2}\, 
    \int\limits_{0}^{q^2}\, {d \xi \over \xi}
    \Bigg\{
    K(a,\eps)
    + 
    G(1,{\bar a}(\xi,\eps),\eps) 
    -
    {1 \over 2}\, 
    \int\limits_{\mu^2}^\xi\, {d \lambda \over \lambda}\
    \gamma({\bar a}(\lambda,\eps))
    \Bigg\} 
 \nn    \\
    &=&
    {1 \over 2}\, 
    \int\limits_{0}^{q^2}\, {d \xi \over \xi}
    \Bigg\{
    G(1,{\bar a}(\xi,\eps),\eps) 
    -
    {1 \over 2}\, 
    \int\limits_{0}^\xi\, {d \lambda \over \lambda}\
    \gamma({\bar a}(\lambda,\eps))
    \Bigg\} 
\end{eqnarray}
where the explicit solution of \eqn{eq:KGdeq}     
for the counterterm function $K$ has been used
\begin{eqnarray}
\label{eq:K-def}
    K(a,\eps) &=&
    - {1 \over 2}\, \int\limits_{0}^{\mu^2}\, {d \lambda \over \lambda}\
    \gamma({\bar a}(\lambda,\eps))
    \, .
\end{eqnarray}
The double poles  $1/\eps^2$            
at each loop order are
generated by the two $\lambda$- and $\xi$-integrations 
over $\gamma$, while the single poles in $\eps$ arise from the outer 
$\xi$-integration over $G$.
Explicit computation, e.g. 
along the lines of refs.~\cite{Bern:2005iz,Moch:2005id}, yields 
\begin{equation}
  \label{eq:phi}
\log \Phi(q^2) \,=\,
\suml   a^\ell
\left[ 
-\frac{\gamma^\pel}{ 4 (\ell \eps)^2} 
-\frac{\cG_0^\pel} {2 \ell \eps} 
- \frac{ \phi^\pel}{2 \ell}
+ \cO(\eps)  
\right]
\left( \mu^2 \over q^2 \right)^{\ell \eps}
\end{equation}
where the boundary condition for $G$ has been chosen as 
\begin{eqnarray}
\label{eq:G-def}
G(1,a,\eps)                     
    &=&   \cG_0(a) 
    + \eps \phi(a) + \cO(\eps^2)  
\end{eqnarray}
which is consistent with \eqn{eq:logPhi-2loop}.
$\cG_0(a)$ is given in \eqn{defG} 
and $\phi(a)$ can be read off from \eqn{eq:logPhi-2loop} as 
\begin{equation}
  \label{eq:defphi}
 \phi(a) =   - \frac{\pi^2 }{6} a +  \cO(a^3)  \, .
\end{equation}
The exponentiation of \eqn{eq:masslessPhi} proceeds trivially with the help of the 
boundary condition for $\Phi$ in $D$ dimensions, i.e., $\Phi(q^2=0) = 1$,
which is implicit also in our choice for $G(1,a,\eps)$ in \eqn{eq:G-def}.
Note also that the all-order result for the form factor in \eqn{eq:masslessPhi}
applies literally to theories with less supersymmetry, e.g., to QCD.
There, the coupling constant ${\bar a}$ has to be read 
as the strong coupling $\alpha_s$ continued to $D$-dimensions and 
the respective QCD expressions for the anomalous dimensions 
$\gamma(a)$ and $\cG_0(a)$ are related to 
\eqns{defgamma}{defG} by the principle of maximal transcendentality.
Moreover, $G(1,a,\eps)$ admits a further decomposition \cite{Dixon:2008gr}
into three terms:
a universal (spin-independent) eikonal anomalous dimension, 
(twice) the coefficient of the $\delta(1-x)$-term in the collinear evolution
kernel, and a process-dependent term accounting for the running 
coupling in the coefficient function of the hard scattering. 
The latter is proportional to the QCD $\beta$-function 
and is, of course, absent in $\cN = 4$ SYM.

We now introduce the wedge function as announced above.
Defining $W(q^2) = \sqrt{\Phi(q^2)}$, we see from \eqn{eq:phi} that
\be
\label{eq:logW}
\log W(q^2) = 
\suml   a^\ell
\left[ 
-\frac{\gamma^\pel}{ 8 (\ell \eps)^2} 
-\frac{\cG_0^\pel} {4 \ell \eps} 
+ w^\pel
+ \cO(\eps) 
\right]
\left( \mu^2 \over  q^2 \right)^{\ell \eps}
\, ,
\ee
where 
\be
\label{defw}
w(a) =  \frac{\pi^2 }{24} a +  \cO(a^3) 
\ee
and $\gamma(a)$ and $\cG_0(a)$ are given in \eqns{defgamma}{defG}.
With \eqn{eq:logW} at our disposal, 
and using exponentiation of the $n$-point amplitude $M_n$ in \eqn{bds},
we can now define the finite remainder $H_n$ via \eqn{finite}
\be
\label{eq:logHn}
\log H_n 
=
{\gamma(a)\over 4}    F_n^\One (\xij) + n \left[ f(a) - w(a)\right]
+ C(a) + R_n (\xij,a) + \cO(\eps)
\ee
thus the IR-divergent pieces of the wedge function 
remove all the IR divergences of the $n$-point amplitude.
At one loop, \eqn{eq:logHn} gives
\be 
\label{HF}
H_n^\One  = F_n^\One -  \frac{n \pi^2}{24}
\ee
allowing us to rewrite it in its final form as
\be
\label{bdsfinite}
\log H_n  =
{\gamma(a)\over 4}    H_n^\One  (\xij) + n D(a) + C(a) + R_n (\xij,a) 
+ \cO(\eps)
\ee
where 
\be
\label{defD}
D(a) = f(a) - w(a) + \frac{\pi^2}{96} \gamma(a) = 
-  \frac{\pi^2 }{ 180} a^2 + \cO(a^3)
\ee
with $C(a)$ given by \eqn{defC}.

Using renormalization group arguments similar to the derivation of 
\eqn{eq:masslessPhi} for the form factor $\Phi$, 
it is obvious that $\log W$, $\log M_n$, and therefore also $\log H_n$ 
in \eqn{bdsfinite}
can be expressed to all orders via (double-)integrals over the respective anomalous dimensions; see e.g. refs.~\cite{Sterman:2002qn,Aybat:2006mz}.

\subsection{Higgs regularization of the form factor}

\begin{figure}[t]
\centerline{
{\epsfxsize3.5cm  \epsfbox{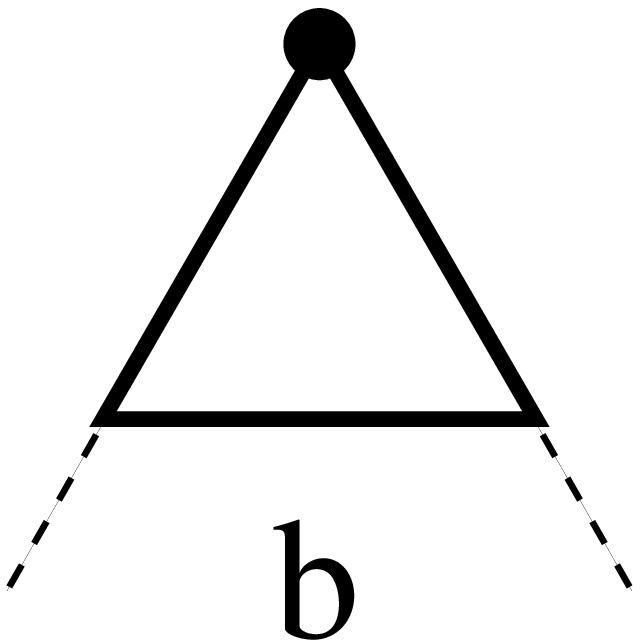}}
{\epsfxsize3.5cm  \epsfbox{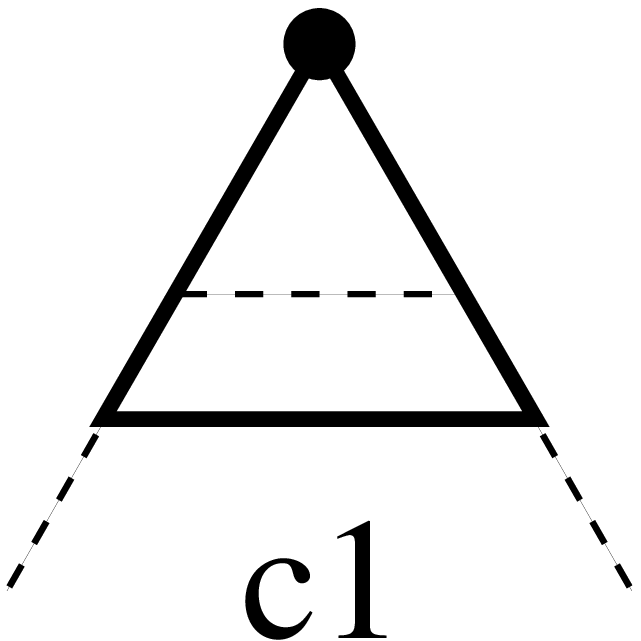}}
{\epsfxsize3.5cm  \epsfbox{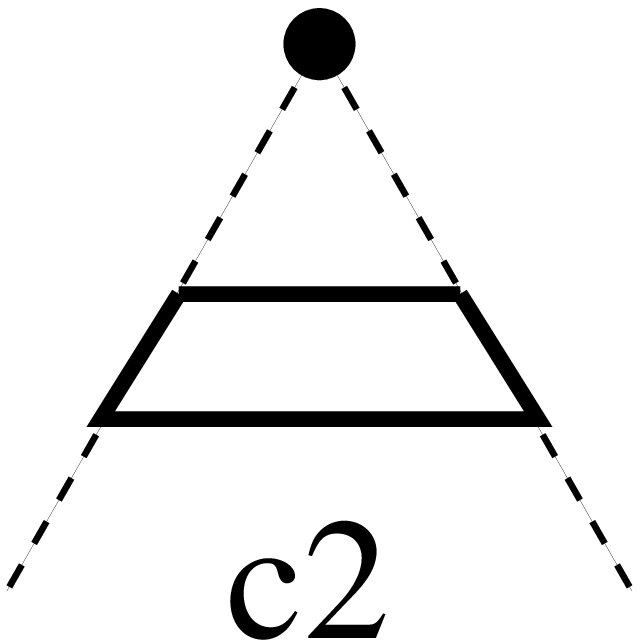}}
{\epsfxsize3.5cm  \epsfbox{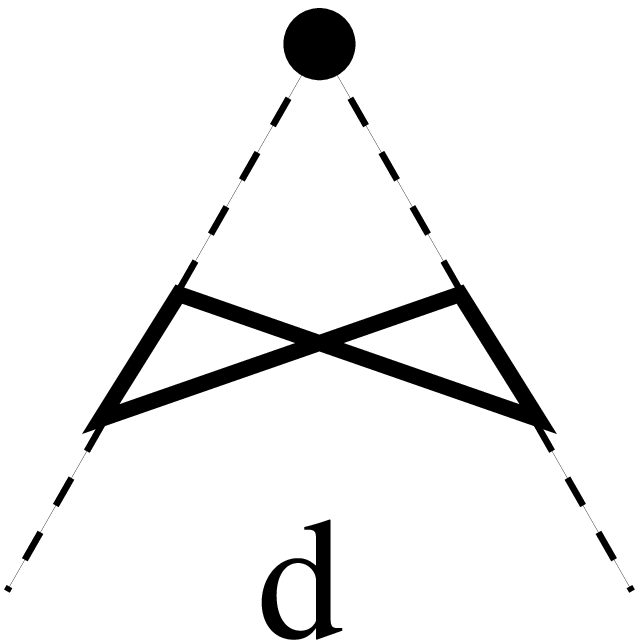}}
}
\caption{One- and two-loop form factor diagrams
for the common-mass Higgs regulator.
The solid/dashed lines 
represent massive/massless adjoint fields.} 
\label{fig-massive}
\end{figure}

\begin{figure}[b]
\centerline{
{\epsfxsize5cm  \epsfbox{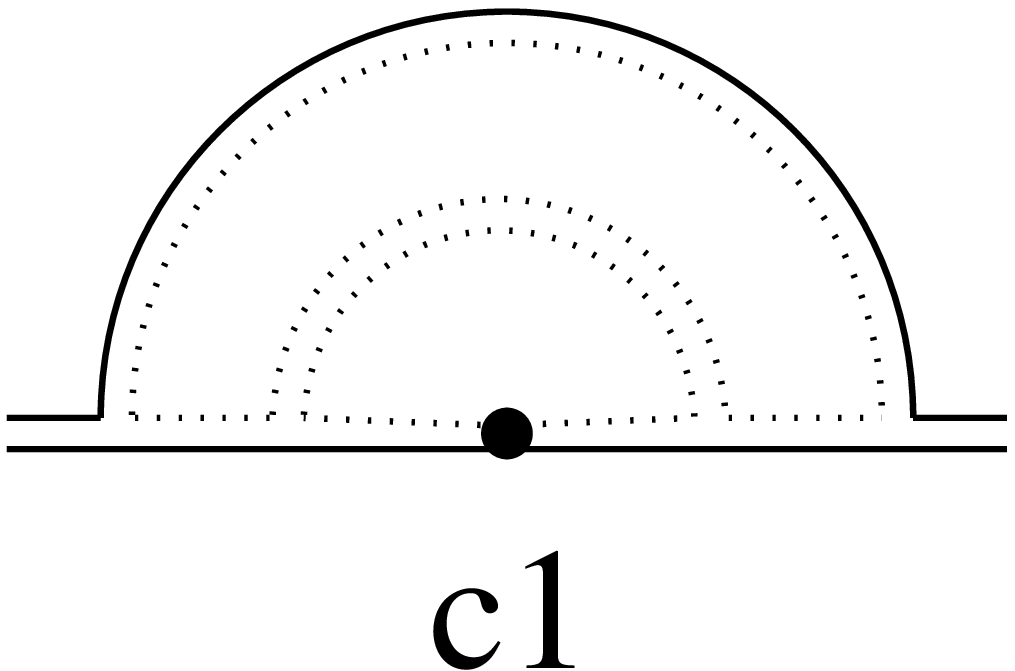}}
{\epsfxsize5cm  \epsfbox{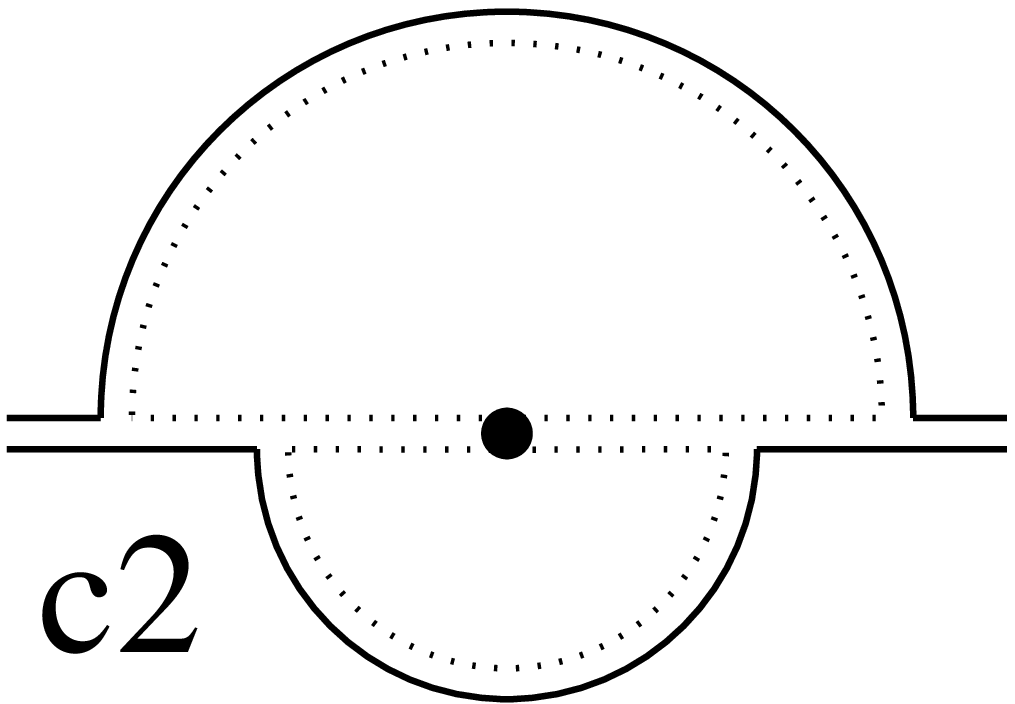}}
{\epsfxsize5cm  \epsfbox{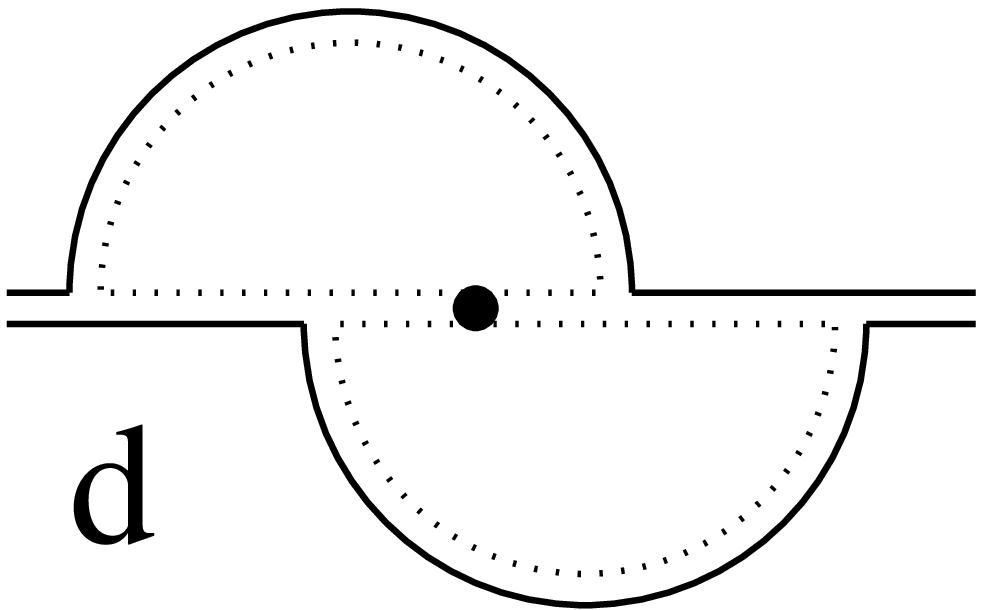}}
}
\caption{Double-line version of the two-loop diagrams 
for the common-mass Higgs regulator. 
The solid/dotted lines represent fundamental fields of $U(M)/U(N)$.}
\label{fig-doubleline}
\end{figure}

Now we turn to the study of the Higgs-regulated $\cN=4$ SYM form factor $\tPhi(q^2)$. 
We will assume that it is given by the same scalar integrals as in fig.~\ref{fig-massless}
except that some of the internal legs are now massive 
\be
\label{deftPhi}
\tPhi (q^2) = 1 
+ a   q^2 \big[-\tI_b \big]
+ a^2 (q^2)^2 
\big[ \half \tI_{c1} + \half  \tI_{c2} + \quarter \tI_d \big] + \cdots
\ee
There are several different mass assignments for the
two-loop integrals (see fig.~\ref{fig-massive}),
as can be seen from the double-line representation
in fig.~\ref{fig-doubleline}.
We have computed these integrals (see appendix \ref{app-integrals})
and they reveal that the Higgs-regulated form factor exponentiates 
to two-loop order
\begin{eqnarray}
  \label{eq:logPhiHiggs-2loop}
\lefteqn{
\log \tPhi(q^2) \,=\, 
}
\\
& &
a \left[ -{1 \over 2} \log^2 \left( \frac{q^2}{m^2} \right)
 + \cO(m^2) \right]
+ 
a^2 
\left[  {\pi^2 \over 12}  \log^2 \left( \frac{q^2}{m^2} \right)
+ \zeta_3  \log \left( \frac{q^2}{m^2} \right)
       +{\pi^2 \over 45} + \cO(m^2) \right]
+ \cO(a^3) \,.
\nn
\end{eqnarray}

The all-loop-order generalization 
of \eqn{eq:logPhiHiggs-2loop} relies on the same 
factorization ansatz discussed before and the separation of scales, 
i.e.  $q^2 \gg m^2$, so that the momentum dependence of $\Phi$ 
is described by the evolution equation~\cite{Collins:1989bt,Mitov:2006xs}
\begin{eqnarray}
  \label{eq:ffmdeq}
  q^2 {\partial \over \partial q^2} \log \Phi(q^2)
  &=&
  {1 \over 2} \: \tK(m^2/\mu^2,a)
+ {1 \over 2} \: \tG(q^2/\mu^2,a)
\end{eqnarray}
and, consistent with \eqn{eq:KGdeq}, 
the renormalization group equation for $\tK$ now reads 
\begin{eqnarray}
  \label{eq:Kmdeq}
  \lim_{m \to 0} \mu {d \over d \mu} \tK(m^2/\mu^2,a) \, = \,
  - \gamma(a) 
\end{eqnarray}
where the limiting procedure $m\to 0$ indicates that we neglect any power
suppressed terms of $\cO(m^2)$.
It is subject to the following solution
\begin{eqnarray}
\label{eq:Km-def}
    \tK(m^2/\mu^2,a) &=&
    \tK(1,a) 
    - {1 \over 2}\, \int\limits_{m^2}^{\mu^2}\, {d \lambda \over \lambda}\
    \gamma(a)
\end{eqnarray}
with a non-vanishing boundary condition $\tK(1,a)$, since the IR sector has been altered 
in contrast to \eqn{eq:K-def}.
The solution to the Higgs-regularized form factor $\tPhi$ then becomes
\begin{eqnarray}
\label{eq:massivePhi}
  \log \tPhi(q^2) 
    &=&
    {1 \over 2}\, 
    \int\limits_{m^2}^{q^2}\, {d \xi \over \xi}
    \Bigg\{
    \tK(1,a) 
    + 
    \tG(1,a)
    -
    {1 \over 2}\, 
    \int\limits_{m^2}^\xi\, {d \lambda \over \lambda}\
    \gamma(a)
    \Bigg\}
 	+  {\tilde\phi}(a)
\end{eqnarray}
where the integration range is naturally cut off in the IR at $m^2$,
i.e. at the mass scale set by the Higgs regulator,
and the function 
\begin{eqnarray}
  \label{eq:phi-finite}
  {\tilde \phi}(a) &=& 
  \frac{\pi^2}{45} a^2 + \cO(a^3)  
\end{eqnarray}
has been introduced to match the fixed-order
computation in \eqn{eq:logPhiHiggs-2loop}.
Note that the evolution equation for the Higgs-regulated 
form factor in $\cN=4$ SYM depends only on the 
sum of $\tK(1,a)$ and $\tG(1,a)$.
To agree with the fixed-order computation in \eqn{eq:logPhiHiggs-2loop},
we choose (cf. \eqn{defGt})
\be
    \tK(1,a) + \tG(1,a) = - 2 {\tilde \cG}_0(a)
\ee
leading to the solution of \eqn{eq:massivePhi} 
\begin{eqnarray}
  \label{eq:phiHiggs}
  \log \tPhi(q^2) &=& 
  - \frac{\gamma(a)}{8} \log^2 \left( \frac{q^2}{m^2} \right)
  - {{\tilde \cG}_0(a)} \log\left( \frac{q^2}{m^2} \right)
  + {\tilde \phi}(a)
  + \cO(m^2)
  \, .
\end{eqnarray}
The exponentiation of \eqn{eq:massivePhi} requires further matching conditions 
for $\tPhi$ to be obtained from explicit $\ell$-loop computations.

A few comments are in order here.
First, matching to fixed-order computations could, in principle, 
also impose the condition 
$\tG = G$, 
i.e. demand that the collinear anomalous dimensions coincide.
This would proceed at the expense of a non-zero result for $\tK(1,a)$.
Next, the Higgs-regulated form factor is finite, 
so that \eqn{eq:massivePhi} can be evaluated in four dimensions.
In theories with broken supersymmetry, e.g. QCD with massive quarks, 
collinear singularities are regulated by the heavy quark masses, 
whereas all soft gluon divergences require dimensional regularization.
In such a case, the analogous functions 
$K$ and $G$ 
have a clear physical interpretation and are independent 
(see e.g. ref.~\cite{Mitov:2006xs}).
For example, the (electric) form factor of a massive quark-anti-quark pair 
in QCD is known to two loops~\cite{Bernreuther:2004ih,Gluza:2009yy}, 
and the analogs of collinear anomalous dimensions
naturally coincide in this case, i.e.  $\tG = G$.

In the case of the common-mass Higgs regulator 
(so that the external states are massless),
the wedge function can again be defined as 
the square root of the form factor (\ref{eq:phiHiggs})
so that
\begin{equation}
  \label{eq:logW-Higgs}
\log \tW (q^2) = 
- \frac{\gamma(a)}{16} 
\log^2 \left( \frac{q^2}{m^2} \right)
- \frac{\tilde{\cal G}_0(a)} {2}
\log\left( \frac{q^2}{m^2} \right)
+\tw(a) 
+ \cO(m^2)
\end{equation}
where  
\be
\label{defwt}
\tw(a) =  \frac{\pi^2 }{90} a^2 +  \cO(a^3) 
\ee
and $\gamma(a)$ and $\tilde{\cG}_0(a)$  were given in \eqns{defgamma}{defGt}.
We now use \eqns{bdshiggs}{eq:logW-Higgs} in \eqn{finite} 
to define the finite part of the $n$-point amplitude 
\ba
\label{eq:logtH}
\log \tH_n 
&=&
\log \tM_n - \sum_{i=1}^n  \log \tW(\xmp )
\nonumber\\
&=&
{\gamma(a)\over 4}    \tF_n^\One  + n \left[ \tf(a) - \tw(a)\right] 
+ \tC(a) + \tR_n  + \cO(m^2) 
\ea
again finding that the IR-divergent pieces of the wedge function 
precisely remove the IR divergences of the $n$-point amplitude.
At one loop, \eqn{eq:logtH} gives
\be
\label{HtFt}
\tH_n^\One  = \tF_n^\One 
\ee
so that we can rewrite the finite part of the amplitude in its final form
\be
\label{bdshiggsfinite}
\log \tH_n =
{\gamma(a)\over 4}    \tH_n^\One  + n \tD(a) + \tC(a) + \tR_n (\xij) + \cO(\eps)
\ee
where 
\be
\label{defDt}
\tD(a) = \tf(a) - \tw(a) = -  \frac{\pi^2 }{ 180} a^2 + \cO(a^3)
\ee
with $\tC(a)$ given by \eqn{defCt}.

In complete analogy to the previous discussion, it is evidently possible to
exploit renormalization group properties
to provide expressions 
for $\log \tW$, $\log \tM_n$, and hence $\log \tH_n$ in \eqn{bdshiggsfinite}
in terms of (double-)integrals over the anomalous dimensions 
similar to \eqn{eq:massivePhi}.

\subsection{Comparison of regulators}

By comparing the results of the last two subsections,
we can see that $\log H_n (\xij) $  as we have defined it in \eqn{finite}
is a good candidate for a regularization-independent 
IR-finite quantity characterizing the planar MHV $n$-point amplitude.
The one-loop hard functions
are identical in both dimensional and Higgs regularization
(cf. eqs.~(\ref{FFt}), (\ref{HF}), and (\ref{HtFt}))
\be
H_n^\One (\xij)  = \tH_n^\One (\xij) \,.
\label{equalH}
\ee
Moreover, calculations through two loops show 
the equality of the kinematic-independent functions appearing in 
the $n$-point amplitude (\ref{bdsfinite}) and (\ref{bdshiggsfinite})
\be
\label{equality}
C(a) = \tC(a),  \qquad D(a) = \tD(a)
\ee
(cf. \eqns{defC}{defCt} and \eqns{defD}{defDt}).
The regulator independence of $C(a)$ was previously observed in ref.~\cite{Henn:2010ir}.
If \eqn{equality} holds to all loops, 
then the regulator independence of the four- and five-point 
hard functions necessarily follows.
For $n=6$ at two loops, 
agreement between the remainder function in dimensional regularization
and massive regularization was observed in ref.~\cite{Drummond:2010mb},
and this agreement is expected for all $n$.
We thus expect the hard function $\log H_n$ to be 
regularization-independent for all $n$-point functions, that is
\be
\log \tH_n (\xij) =\log H_n (\xij)
\ee
for dimensional and Higgs regularizations.

\section{Differential-mass Higgs regulator} 
\setcounter{equation}{0}
\label{sect-diff} 

In section \ref{sect-finite}, we defined an IR-finite hard function 
$\log H_n$ for the $n$-point amplitude,
and showed (through two loops)
that it has the same form (including constants) for 
dimensional regularization and for a common-mass Higgs regulator.
In this section, we generalize our discussion to a more general class of regulators,
\viz, the Higgs regulator with arbitrary distinct masses.
This is also interesting from the point of view of collider phenomenology.
In QCD, amplitudes with different masses have been considered to two loops 
for the heavy-to-light transitions,
i.e., the (axial)-vector form factor with one massive 
and one massless 
quark~\cite{Bonciani:2008wf,Asatrian:2008uk,Beneke:2008ei,Bell:2008ws}.
Also, electroweak logarithms in four-fermion processes at high energy 
arising from loop corrections with massive $W$- and $Z$-gauge bosons 
have been considered to two loops (see e.g. ref.~\cite{Jantzen:2005az}).

Recall that breaking the $U(N+M)$ symmetry of $\cN=4$ SYM theory
to $U(N)\times  U(1)^M$ by assigning distinct 
vacuum expectation values to one of the scalar fields 
results in non-zero       
masses $|m_i-m_j| $       
for the off-diagonal adjoint fields
and distinct masses $m_i$ for the internal propagators of the scalar integrals
that characterize loop amplitudes.
(In fact, extended dual conformal invariance requires 
the freedom to vary the masses.)
One can then define a differential-mass Higgs regulator by taking
all the masses $m_i$ to zero.
More precisely,  if $m_i = \alpha_i m$,
the ``uniform small mass limit''
is defined as the limit $m\to 0$ with $\alpha_i$ held fixed.
Regulator independence means that the result does
not depend on the choice of $\alpha_i$.

Because the external legs of the $n$-point amplitude 
now have distinct masses  $|m_i-m_{i+1}|$,
it is no longer possible to define the wedge function
$\wedge (\xmp)$ as the square root of a form factor
as we did in sec.~\ref{sect-finite}. 
In fact, it is not obvious what the 
operational definition of $\wedge(\xmp)$  
for the differential-mass Higgs regulator should be,  
and we leave this question to the future.   

For now,  we adopt a different 
approach by decomposing the one-loop $n$-point amplitude into 
an IR-divergent and a manifestly regulator-independent IR-finite piece, 
and then defining the one-loop wedge function in terms of the former. 
The extended dual conformal invariance of the BDS ansatz 
then allows us to generalize this to an all-loop wedge function.

\subsection{One-loop amplitude with differential-mass Higgs regulator} 
\label{subsect-integral-decomposition}

\begin{figure}[t]
\centerline{
{\epsfxsize4cm  \epsfbox{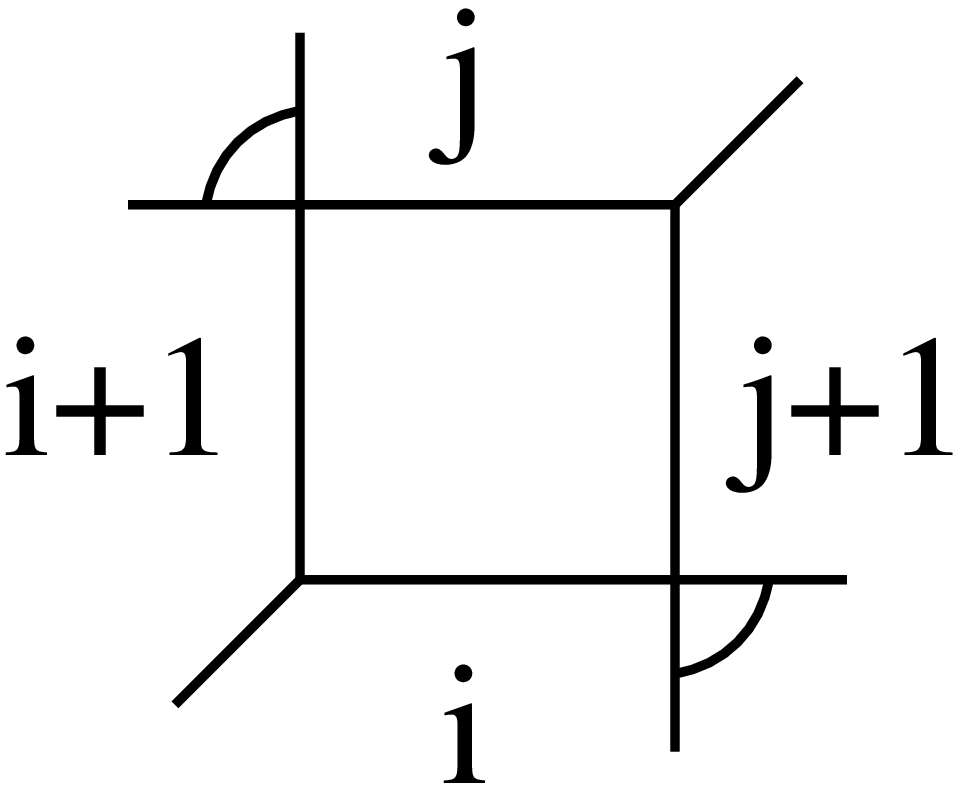}}
}
\caption{Two-mass-easy diagram corresponding to the integral $\twome$.}
\label{fig-2me}
\end{figure}

As is well known,
the one-loop MHV $n$-point amplitude in dimensional regularization can
be written as a sum of two-mass-easy (and one-mass\footnote{The one-mass 
integrals are just the special cases $I^{\rm 2me}_{i,i+2}$.}) 
scalar box integrals
\cite{Bern:1994cg,Bern:1994zx}
\be
\label{sumoftwomasseasy}
M_n^\One =  - {1 \over 8} \sum_{i=1}^n  \sum_{j=i+2}^{i+n-2} \twome  \,.
\ee
We will assume that the amplitude on the Coulomb branch is given,
at least up to $\cO(m^2)$, 
by the same set of integrals, 
with the mass configuration dictated by dual conformal symmetry.
The two-mass-easy diagram in fig.~\ref{fig-2me} corresponds to 
the integral
\be
\label{2me}
\twome = \int  {d^4 x_0   \over i \pi^2}
{
 \hx_{ij}^2 \hx_{i+1,j+1}^2 -\hx_{i+1,j}^2 \hx_{i,j+1}^2
\over
\hx_{0i}^2 \hx_{0,i+1}^2 \hx_{0j}^2  \hx_{0,j+1}^2 }
\ee
where 
$\hx_{ij}^2 =  \xij + (m_i - m_j)^2$
with $m_0 = 0$.
Later, we will take the uniform small mass limit,
so henceforth we drop all mass dependence from the numerators, 
as those terms would only contribute at $\cO (m^2)$.

It is known that one can decompose \eqn{2me} into a sum 
of IR-divergent triangle integrals and 
an IR-finite six-dimensional integral
(see e.g. ref.~\cite{Dixon:2011ng})
\begin{eqnarray}
\twome 
= 
\twome  \big|_{\rm tri} 
+
\twomesix 
+ \cO(m^2)
\end{eqnarray}
where
\ba
\twome  \big|_{\rm tri}
&=& \int \frac{d^{4}x_{0}}{i \pi^2} 
\Bigg[
{ 
{x}_{i+1,j+1}^2 - {x}_{i+1,j}^2   
\over
{\hx_{0,i+1}^2 \hx_{0,j}^2 \hx_{0,j+1}^2 }
}
{+}{ 
{x}_{i,j}^2 - {x}_{i,j+1}^2   
\over
{\hx_{0,i}^2 \hx_{0,j}^2 \hx_{0,j+1}^2 }
}
{+}{ 
{x}_{i+1,j+1}^2 - {x}_{i,j+1}^2   
\over
{\hx_{0,i}^2 \hx_{0,i+1}^2 \hx_{0,j+1}^2 }
}
{+}{ 
{x}_{i,j}^2 - {x}_{i+1,j}^2   
\over
{\hx_{0,i}^2 \hx_{0,i+1}^2 \hx_{0,j}^2 }
}
\Bigg]
\label{tri}
\ea
and 
\begin{eqnarray}
\label{2me6D}
\twomesix 
= \frac{2}{x_{ab}^2}  \int  {d^6 x_0   \over i \pi^3}
{
 x_{ij}^2 x_{i+1,j+1}^2 -x_{i+1,j}^2 x_{i,j+1}^2 
\over
x_{0i}^2 x_{0,i+1}^2 x_{0j}^2  x_{0,j+1}^2 }
\end{eqnarray}
with $x_a$ and $x_b$ the two solutions of the equations
$x_{0i}^2 = x_{0,i+1}^2 = x_{0j}^2  = x_{0,j+1}^2 =0$.
In appendix \ref{app-decomp} of this paper,
we review the derivation of this decomposition using 
twistor methods.
Since the six-dimensional integral (\ref{2me6D}) is IR-finite,
it is independent of which IR-regulator
we employ to regulate the $n$-point amplitude. 
Therefore in the decomposition of the differential-mass Higgs-regulated
amplitude
\be
\label{Mn1wedges}
M_n^\One =
\sum_{i=1}^n   \wdiff^\One  (\xmp ) + H_n^\One  (\xij) + \cO(m^2)
\ee
a natural candidate for the regulator-independent hard function is
\be
H_n^\One =  - {1 \over 8} \sum_{i=1}^n  \sum_{j=i+2}^{i+n-2} \twomesix  \,.
\ee
Moreover, the sum of the one-loop wedge functions in \eqn{Mn1wedges} 
will then be given by 
the sum of triangle diagrams (\ref{tri})
that contribute to the $n$-point amplitude.
Most of the triangle diagrams in this sum cancel, leaving 
\ba
\sum_{i=1}^n   \wdiff^\One  (\xmp ) 
&=&
- \frac{1}{2} 
 \sum_{i=1}^n  
\int  {d^4 x_0 \over i\pi^2}
{ 
\xmp
\over
\hx_{0,i-1}^2 \hx_{0i}^2 \hx_{0,i+1}^2 }\,.
\ea
This suggests the following one-loop expression for the wedge 
function\footnote{The subscripts
on the wedge function refer to its dependence on 
$m_{i-1}$, $m_i$, and $m_{i+1}$.}
\be
\label{wedge}
\wdiff^\One (\xmp ) = 
- \frac{1}{2} 
 \int  {d^4x_0 \over i \pi^2}  
\frac{\xmp}{ \hx_{0,i-1}^2 \hx_{0i}^2 \hx_{0,i+1}^2 }
\ee
although we could in principle add a contribution that vanishes upon 
summing over $i$.
\Eqn{wedge} reduces to our previous definition
$W(q^2) = \sqrt{\Phi(q^2)}$
when $m_{i-1} = m_i = m_{i+1}$.
Evaluating \eqn{wedge} in the uniform small mass limit,
we obtain
\be
\label{oneloopwedge}
\wdiff^\One (\xmp ) 
=
- \frac{1}{4} \log^2 \left( x_{i-1,i+1}^2 \over m_i^2 \right) 
-  \Li_2 \left( 1 - \frac{m_{i-1}}{m_i} \right)
-  \Li_2 \left( 1 - \frac{m_{i+1}}{m_i} \right)
+ \cO(m^2)\,.
\ee
Substituting \eqn{oneloopwedge} into \eqn{Mn1wedges}
and using the identity $ \Li_2 (1-z) + \Li_2 (1- z^{-1}) + \frac{1}{2} \log^2 z = 0$,
we finally obtain the differential-mass Higgs-regulated 
$n$-point amplitude 
\ba
\label{decomp}
M_n^\One 
&=&
- \frac{1}{4} \sum_{i=1}^n
\log^2 \left( x_{i-1,i+1}^2 \over m_i^2 \right) 
+ \frac{1}{8} \sum_{i=1}^n
\log^2 \left( m_{i+1}^2 \over m_i^2 \right) 
+ H_n^\One  (\xij)
+ \cO(m^2)
\ea
where, as discussed above, $H_n^\One (\xij)$ 
is IR-finite and regulator-independent;
in particular, it does not depend on $\alpha_i$ in the uniform small mass limit
$m\to 0$, where $m_i =  \alpha_i m $ with $\alpha_i$ fixed.

Note that although $M_n^\One$ has extended dual conformal invariance,
the decomposition 
into $H^\One_{n}$ and the specific IR-divergent pieces in \eqn{decomp}
breaks this symmetry.
This is not surprising, as triangle integrals manifestly violate
dual conformal symmetry.

\subsection{Higher loops}

In the previous subsection, we
derived an expression (\ref{oneloopwedge}) for the one-loop wedge function
valid for the differential-mass Higgs regulator.
We now use the extended dual conformal invariance of the
amplitude and the BDS ansatz to derive the 
explicit form of the wedge function at higher loops.

Recall that extended dual conformal invariance implies 
\cite{Alday:2009zm} that the amplitude can only be a function of 
\be
\label{uij}
\uij = {m_i m_j \over \xij + (m_i-m_j)^2 } \,.
\ee
For the common-mass Higgs regulator, this reduces to $\uij = m^2/ \xij$.
Hence, assuming the validity of the
all-loop expression (\ref{bdshiggs}) for 
the common-mass Higgs-regulated amplitude,
its unique generalization is obtained by replacing 
$\xij$ with $m^2/\uij$ everywhere to obtain
\ba
\log \tM_n &=& \sum_{i=1}^n
\left[ - \frac{\gamma(a)}{16} 
\log^2 \left( \frac{1}{\ump}\right)
- \frac{\tilde{\cal G}_0(a)} {2}
\log\left( \frac{1}{\ump}\right)
\right]
\\
&& \,+\, \frac{\gamma(a)}{4} \, H_n^{(1)}  (m^2/\uij) 
\,+\, n \tf (a)
\,+\, \tC (a)
\,+\, \tR_n  (m^2/\uij, a)
\,+\, \cO(m^2) 
\nn
\ea
where we have also used \eqns{HtFt}{equalH}.
In the uniform small mass limit, 
we can neglect $(m_i-m_j)^2$ relative to $\xij$
in \eqn{uij} so that $\uij$ becomes  $m_i m_j/\xij$
yielding 
\ba
\label{bdsdiff}
\log \tM_n &=& \sum_{i=1}^n
\left[ - \frac{\gamma(a)}{16} 
\log^2 \left( \frac{\xmp } {m_{i-1} m_{i+1}} \right)
- \frac{\tilde{\cal G}_0(a)} {2}
\log\left( \frac{\xmp } {m_{i-1} m_{i+1}} \right)
\right]
\\
&& \,+\, \frac{\gamma(a)}{4} \, 
H_n^\One  \left( m^2 \xij \over m_i m_j \right)
\,+\, n \tf (a)
\,+\, \tC (a)
\,+\, \tR_n  \left( {m^2 \xij \over m_i m_j}, a\right)
\,+\, \cO(m^2) \,.
\nn
\ea
The apparent dependence of $\tR_n$ on $m_i$
is illusory since the mass dependence cancels out
in the dual conformal cross ratios on which 
$\tR_n$ only depends,
so that 
$\tR_n  ( m^2 \xij /m_i m_j, a) = \tR_n  ( \xij , a)$.
There does, however, remain some dependence on $m_i$ in $H^\One_n$.

Applying the same reasoning as above to the one-loop
amplitude, we obtain
\be
\label{diff}
M_n^\One 
= - \frac{1}{4} \sum_{i=1}^n
\log^2 \left( \xmp \over m_{i-1} m_{i+1}  \right) 
+ H_n^\One  \left( \frac{m^2 \xij}{m_i m_j} \right)
+ \cO(m^2)  \,.
\ee
Now since \eqns{decomp}{diff} are both valid expressions for the
differentially-regulated one-loop amplitude, we deduce that
\ba
\label{onelooprln}
H_n^\One  \left( \frac{m^2 \xij}{m_i m_j} \right)
&=& 
 H_n^\One  \left( \xij \right) 
+\frac{1}{4} \sum_{i=1}^n
\left[ 
   \log^2 \left( x_{i-1,i+1}^2 \over m_{i-1} m_{i+1}  \right) 
- \log^2 \left( x_{i-1,i+1}^2 \over m_i^2 \right) 
\right]
\nn\\
&&
+ \frac{1}{8} \sum_{i=1}^n
\log^2 \left( m_{i+1}^2 \over m_i^2 \right) 
+ \cO(m^2) \,.
\ea
Substituting  \eqn{onelooprln} into \eqn{bdsdiff}
and using \eqn{defDt}, we
obtain\footnote{The  apparent difference between the $\tilde{\cG}_0(a)$
terms disappears on performing the sum over $i$.}
\ba
\label{bdsdifftwo}
\log \tM_n 
&=&
\sum_{i=1}^n
\left[ - \frac{\gamma(a)}{16} 
\log^2 \left( \frac{\xmp }{m_i^2} \right)
- \frac{\tilde{\cal G}_0(a)} {2}
\log\left( \frac{\xmp }{m_i^2} \right)
+ \frac{\gamma(a)}{32} 
\log^2 \left( m_{i+1}^2 \over m_i^2 \right) 
+ \tw(a) \right]
\nonumber\\
&& \,+\, \frac{\gamma(a)}{4} \, H_n^{(1)}  ( \xij )
\,+\, n \tD (a)
\,+\, \tC (a)
\,+\, \tR_n  (\xij, a) 
\,+\, \cO(m^2)
\ea
where now only the terms in the sum on the first line depend on the regulator,
while the pieces on the second line are all regulator-independent.    
Recalling that 
\be
\log M_n =  \sum_{i=1}^n  \log \wdiff (\xmp ) + \log H_n
\label{decom}
\ee
we deduce from \eqn{bdsdifftwo} that the
all-order expression for the sum of wedge functions
in differential-mass Higgs regularization is\footnote{
\Eqn{sumwedge} does not allow us, however, to identify the
individual terms of the sum over $i$.} 
\ba
\label{sumwedge}
&&\sum_{i=1}^n  \log \wdiff (\xmp ) 
\\
&& 
\hskip1cm
=\sum_{i=1}^n
\left[ - \frac{\gamma(a)}{16} 
\log^2 \left( \frac{\xmp }{m_i^2} \right)
- \frac{\tilde{\cal G}_0(a)} {2}
\log\left( \frac{\xmp }{m_i^2} \right)
+ \frac{\gamma(a)}{32} 
\log^2 \left( m_{i+1}^2 \over m_i^2 \right) 
+ \tw(a) 
\right]
\nn
\ea
and the regulator-independent IR-finite  piece is, as before
\be
\log H_n  =
{\gamma(a)\over 4}    H_n^\One  (\xij) + n D(a) + C(a) + R_n (\xij,a) 
\ee
where we have dropped all the tildes.

We observe again that, 
although the amplitude $\log M_n$ has extended dual 
conformal invariance,
the separate terms in the decomposition (\ref{decom})
do not.

\subsection{Relation to anomalous dual conformal Ward identity} 
\setcounter{equation}{0}
\label{sect-anom}

In the previous section, we obtained the all-loop expression
(\ref{sumwedge}) for the sum of wedge functions
by assuming \eqn{bdshiggs}.
In this section, 
we show inversely that \eqn{bdshiggs} follows from \eqn{sumwedge}. 
Therefore, it would be interesting to have a first-principles derivation 
of \eqn{sumwedge}.

The $n$-point amplitude has exact extended dual conformal symmetry, 
and so is annihilated by the generator of dual special conformal transformations
\be
\hat{K}^{\mu} \log M_n 
\equiv
\sum_{i=1}^{n}  \left[  2 x_{i}{}^{\mu} \left( x_{i}^{\nu}
\frac{\partial}{\partial x_{i}^{\nu}} + m_{i} \frac{\partial}{\partial
m_{i}} \right) - (x_{i}^2 +m_{i}^2) \frac{\partial}{\partial x_{i\mu}}
\right] \log M_n =0\,.
\label{exact}
\ee
In ref.~\cite{Alday:2009zm},
it was suggested that the IR-divergent properties of 
the Higgs-regulated amplitude provides a relation
between this exact Ward identity 
and the anomalous dual conformal Ward identity
for the IR-finite part of the $n$-point amplitude
that was originally derived in a Wilson loop context \cite{Drummond:2007cf,Drummond:2007au}.
We will see that this is indeed the case. 

As we have seen, the $n$-point amplitude can be written as
\be
\log M_n =  \sum_{i=1}^n  \log \wdiff (\xmp ) + \log H_n \,.
\ee
Using the expression (\ref{sumwedge})
for the sum of the wedge functions, 
one can easily show that
\be
\hat{K}^{\mu} \sum_{i=1}^n  \log \wdiff (\xmp ) 
=
-\frac{\gamma(a) }{4} \sum_{i=1}^{n}  \left[ x_{i-1}^\mu - 2 x_i^\mu + x_{i+1}^\mu \right] \log \left(\xmp \right) 
\ee
which by virtue of \eqn{exact} implies that 
\be
\hat{K}^{\mu} \log H_n
=
\frac{\gamma(a) }{4} \sum_{i=1}^{n}  \left[ x_{i-1}^\mu - 2 x_i^\mu +
x_{i+1}^\mu \right] \log \left(\xmp \right) \,.
\ee
But $\log H_n$ is regulator-independent,
i.e., has no dependence on $m_i$ in the uniform small mass limit, so 
the $m$-dependent pieces in $ \hat{K}^{\mu}$ drop out when
acting on $\log H_n$ 
and we have
\begin{equation}
{K}^{\mu} \log H_n \equiv
\sum_{i=1}^{n}  \left[  2 x_{i}{}^{\mu}  x_{i}^{\nu}
\frac{\partial}{\partial x_{i}^{\nu}} - x_{i}^2 \frac{\partial}{\partial
x_{i\mu}} \right] \log H_n
= \frac{\gamma(a) }{4} \sum_{i=1}^{n}  \left[
x_{i,i+1}^{\mu} \, \log \frac{x^2_{i,i+2}}{x^2_{i-1,i+1}} \right] 
\label{anom}
\end{equation}
which is precisely the anomalous dual conformal Ward identity \cite{Drummond:2007cf,Drummond:2007au}.
This in turn implies \eqn{bdshiggs}.

We thus see that the decomposition of the amplitude 
into contributions which separately do not possess extended 
dual conformal invariance was necessary to obtain the anomalous dual 
conformal Ward identity for the finite (regulator-independent)
part of the amplitude.

\section{Discussion} 
\setcounter{equation}{0}
\label{sect-discussion} 

In this paper, we have given a prescription for 
defining an unambiguous, regulator-independent 
IR-finite part of the MHV $n$-point scattering amplitude
in planar $\cN=4$ SYM theory.
This prescription involves the definition of an
IR-divergent wedge function associated with 
a pair of adjacent external legs of the amplitude.
The IR-finite part of the amplitude is then defined
as the quotient of the $n-$point amplitude 
by the product of wedge functions, 
cf. \eqn{log-hard-function}.

For regulators that leave the external legs massless
(e.g., dimensional regularization or the common-mass
Higgs regulator), the wedge function can be naturally 
defined in terms of a form factor $\Phi$ which has 
the same IR-divergences.   Computation of this
form factor in dimensional regularization and in 
the common-mass Higgs regularization through two loops
shows that the IR-finite part of the amplitude is identical
for these two regularizations.
For the more general differential-mass Higgs regulator,
which gives (small) masses to the external legs,
a wedge function that results
in a regulator-independent hard function
can still be calculated, 
but an operator definition in this case is still
lacking.

We remark that the idea of defining a 
regulator-independent finite hard function 
can also be applied to other objects, 
e.g. Wilson loops. 
This is particularly interesting in the 
context of the Wilson loop/scattering amplitudes duality, 
since the two objects have different types of divergences, 
viz., UV and IR divergences respectively.
Although these divergences are related, 
defining hard functions for both objects could be useful 
for stating the duality in a regulator-independent way.

There exist in the literature other procedures for 
removing the IR divergences of scattering amplitudes.
For example, for non-MHV amplitudes, one can define 
an IR-finite ratio function \cite{Drummond:2008vq} 
by factoring out the entire MHV amplitude, 
using the universality of IR divergences, i.e. 
that they do not depend on the helicity configuration.
Another example involves MHV amplitudes with $n\ge 6$ external legs.
Since the four- and five-point amplitudes 
(or, equivalently, Wilson loops) are known up to kinematic-independent 
functions, they can be used to remove the divergences of 
higher-point amplitudes by defining suitable ratios \cite{Alday:2010ku}.
This latter procedure preserves dual conformal symmetry.

The hard functions defined in this paper are not dual conformal invariant; 
they have the advantage, however, of allowing us to study 
the $n=4$ and $n=5$ cases as well. 
In particular, it would be interesting to understand better the
systematics of how the BES equation \cite{Beisert:2006ez} for $\gamma(a)$ 
arises from the loop expansion of the four-point amplitude.

The breaking of dual conformal invariance by the hard function
also implies an intimate connection between 
the anomalous dual conformal Ward identity it satisfies 
and the IR divergences (wedge functions) of 
differential-mass regulated amplitudes.
A first-principles derivation of the latter would therefore
be most interesting.

Finally, we believe that, although our investigation has been specialized
to $\cN=4$ SYM theory, the insight into the interplay between regulator,
kinematics, and soft and collinear momentum configurations applies to
many gauge field theories, including those with broken supersymmetry, 
such as QCD,
and also to electroweak radiative corrections in the Standard Model.

\section*{Acknowledgments}
It is a pleasure to thank Z.~Bern, J.~Drummond, and G.~Korchemsky
for interesting discussions.
All of the authors also wish to express their gratitude to 
the Kavli Institute for Theoretical Physics,
where part of this work was carried out during the
``The Harmony of Scattering Amplitudes'' program in spring 2011.
This research was supported in part by 
the European-Union funded network {\it LHCPhenoNet} contract 
No. PITN-GA-2010-264564, 
as well as the National Science Foundation under Grant Nos. 
PHY05-51164 and PHY07-56518.

\appendix
\section{Results for one- and two-loop integrals}
\setcounter{equation}{0}
\label{app-integrals} 

In this appendix, we list the results for various 
massless and massive three-point integrals that contribute
to the form factors computed in this paper.
We use the 
mostly-plus     
metric,
the propagators are of the form $k^2 + m^2$,
and the measure of each internal loop momentum 
is multiplied by a factor of 
$(\mu^2 \e^{-\gamma})^\eps/(i \pi^{d/2} )$. 
The massless integrals shown in fig.~\ref{fig-massless}
are dimensionally-regulated, giving rise to 
the following Laurent 
expansions \cite{Gonsalves:1983nq,vanNeerven:1985xr,Smirnov:2006ry}
\ba
I_b
&=&  
(\mu^2 \e^{-\gamma})^\eps
\int  { d^d x_0 \over i \pi^{d/2} }
{1 \over x_{01}^2  x_{02}^2 x_{03}^2 }
=
{1 \over q^2} \left( \mu^2 \over q^2\right)^\eps 
\left[   {1 \over \eps^2} - {\pi^2 \over 12} - {7 \zeta_3 \over 3}\eps  
- {47 \pi^4 \over 1440}\eps^2  + \cO(\eps^3)  
\right] \,,\\
I_c 
&=& 
{1 \over (q^2)^2} \left( \mu^2 \over q^2\right)^{2\eps }
\left[ {1 \over 4 \eps^4} + {5 \pi^2 \over 24 \eps^2}
+ {29 \zeta_3 \over 6 \eps} + {3 \pi^4 \over 32} + \cO(\eps) \right]
\,,\\
I_d 
&=& 
{1 \over (q^2)^2} \left( \mu^2 \over q^2\right)^{2\eps }
\left[ {1 \over \eps^4} - { \pi^2 \over\eps^2}
- {83 \zeta_3 \over 3 \eps} - {59 \pi^4 \over 120} + \cO(\eps) \right]  \,,
\ea
where $q^2 = x_{13}^2$.
The integrals shown in fig.~\ref{fig-massive}
use a common-mass Higgs regulator,
and can be evaluated to give 
\ba
\tI_b &=&
\int  { d^4 x_0 \over i \pi^2}
{1 \over (x_{01}^2 +m^2) (x_{02}^2 +m^2) (x_{03}^2 +m^2) }
=
{1 \over q^2} 
\left[  
{1 \over 2} \log^2\left( q^2 \over m^2 \right) + \cO(m^2) 
\right]
\,,\\
\tI_{c1} &=&
{1 \over (q^2)^2} 
\left[
  {1 \over 24} \log^4\left( q^2 \over m^2 \right) 
+ {\pi^2 \over 3} \log^2\left( q^2 \over m^2 \right) 
- 8 \zeta_3 \log\left( q^2 \over m^2 \right) 
+ {\pi^4 \over 10} + \cO(m^2) 
\right]
\,,\\
\tI_{c2} &=&
{1 \over (q^2)^2} 
\left[ 
  {1 \over 24} \log^4\left( q^2 \over m^2 \right) 
+ {\pi^2 \over 3} \log^2\left( q^2 \over m^2 \right) 
- 10 \zeta_3 \log\left( q^2 \over m^2 \right) 
+ {13 \pi^4 \over 60} + \cO(m^2) 
\right]
\,,\\
\tI_d &=&
{1 \over (q^2)^2} 
\left[
  {1 \over 3} \log^4\left( q^2 \over m^2 \right) 
- \pi^2 \log^2\left( q^2 \over m^2 \right) 
+ 40\zeta_3 \log\left( q^2 \over m^2 \right) 
- {49\pi^4 \over 90} + \cO(m^2) 
\right] \,.
\ea

\section{Decomposition of the 2me box integral} 
\setcounter{equation}{0}
\label{app-decomp} 

In this appendix, we derive the decomposition of the
2me box integral into triangle integrals and
an IR-finite six-dimensional integral 
that was used in sec.~\ref{subsect-integral-decomposition}.

We begin by rewriting the dual conformal invariant integral 
of equation (\ref{2me})
in terms of momentum twistors \cite{Hodges:2009hk};  
see ref.~\cite{ArkaniHamed:2010gh} for a pedagogical introduction 
to this topic.
A point $x_i$ in dual space corresponds to a
(projective) line $Z^A_{i-1} Z^B_{i}$ in momentum twistor space.
The invariant $\xij$ can be expressed  as
\be
\label{fourbracket}
\xij = {\vev{i-1,i,j-1,j} \over \vev{i-1,i} \vev{j-1,j} }
\ee
where the twistor four-bracket is
\be
\vev{ \al,\bet,\gamma,\delta}
= \eps_{ABCD}  Z^A_\al Z^B_\bet Z^C_\gamma Z^D_\delta \,.
\ee
We introduce the infinity bitwistor $I^{AB}$, which when 
contracted with $ Z^C_\al Z^D_\bet $ gives the two-bracket
\be
\vev{ \al,\bet} = \eps_{ABCD}  I^{AB} Z^C_\al Z^D_\bet  \,.
\ee
Finally, we introduce a modified mass-regulated four-bracket 
\be
\label{modified}
\vev{ \al,\bet,\gamma,\delta}_i 
\equiv
\vev{ \al,\bet,\gamma,\delta} + m_i^2 \vev{ \al,\bet} \vev{\gamma,\delta} \,.
\ee
Rewriting \eqn{2me} using \eqns{fourbracket}{modified}, we obtain
\be
\label{2metwistor}
\twome = \int \dZ
{ N_{ij}
\over
\vev{ \al,\bet, i-1,i }_i \vev{ \al,\bet, i, i-1 }_{i+1}
\vev{ \al,\bet, j-1,j }_j \vev{ \al,\bet, j, j-1 }_{j+1}
}
+\cO(m^2)
\ee
where
\be
N_{ij} \equiv
\vev{i-1,i,j-1,j} \vev{i,i+1,j,j+1} -\vev{i,i+1,j-1,j} \vev{i-1,i,j,j+1}
\ee
and where $d^4 x_0 \to d^4 Z_{\al\bet}/ \vev{\al,\bet}^4$.

We now decompose the two-mass-easy integral into 
a sum of IR-divergent and IR-finite contributions using a twistor identity.
To derive this identity, consider the infinity bitwistor 
$I^{AB}$ and expand it in the basis spanned by the six simple bitwistors
\be
\label{infinity}
I^{AB}  =  c_{i-1,i}  Z^A_{i-1}  Z^B_i
        +  c_{i,i+1}  Z^A_{i}  Z^B_{i+1}
        +  c_{j-1,j}  Z^A_{j-1}  Z^B_j
        +  c_{j,j+1}  Z^A_{j}  Z^B_{j+1}
        +  c_{i,j}  Z^A_{i}  Z^B_j
        +  c_{\bi,\bj}  Z^A_{\bi}  Z^B_{\bj}
\ee
where
$ Z^A_{\bi}  Z^B_{\bj}$
denotes the line in momentum twistor space formed by the intersection of 
\break
$(i-1,i,i+1)$ and $(j-1,j,j+1)$.
Contracting \eqn{infinity} with $Z^C_\al Z^D_\bet$, we obtain
the identity 
\ba
\label{iden}
\vev{\al\bet} &=& c_{i-1,i}  \vev{\al,\bet, i-1,i}
        +  c_{i,i+1}  \vev{\al,\bet, i,i+1}
        +  c_{j-1,j}  \vev{\al,\bet, j-1,j}
\nn\\
&&      + ~ c_{j,j+1}  \vev{\al,\bet, j,j+1}
        +  c_{i,j}  \vev{\al,\bet, ij} 
        +  c_{\bi,\bj}  \vev{\al,\bet,\bi\bj} \,.
\ea
We multiply and divide the integrand of \eqn{2metwistor} 
by $\vev{\al,\bet}$,
using \eqn{iden} to rewrite the numerator.
The first four terms will each cancel one of the propagators\footnote{We rewrite
the numerator term $\vev{\al,\bet, i-1,i}$ as 
$\vev{\al,\bet, i-1,i}_i - m_i^2 \vev{\al\bet} \vev{i-1,i}$
and then drop the $\cO(m^2) $ piece.}
resulting in four triangle integrals, 
whereas the last two terms remain box integrals
\be
\twome  = 
\twome  \big|_{\rm tri}
+\twome  \big|_{\rm box} 
+\cO(m^2) \,.
\ee
The coefficients $c_{i-1,i}$, etc. 
may be determined by contracting \eqn{infinity}
with each of the basis elements in turn, e.g., 
\be
\vev{ i-1,i } = 
c_{j-1,j}  \vev{ i-1,i,j-1,j } +  c_{j,j+1} \vev{ i-1,i, j,j+1 }
\ee
and five other equations.
Thus we find 
\ba
\twome  \big|_{\rm tri}
&=& \int \dZ {1 \over \vev{\al,\bet} }
\Bigg[
{ 
\vev{j-1,j} \vev{i,i+1,j,j+1} -\vev{j,j+1} \vev{i,i+1,j-1,j} 
\over
\vev{ \al,\bet, i, i+1 }_{i+1} \vev{ \al,\bet, j-1,j }_j \vev{ \al,\bet, j, j+1 }_{j+1}
}
\nn\\
&&\hskip 3mm
+{ 
\vev{j,j+1} \vev{i-1,i,j-1,j}- \vev{j-1,j} \vev{i-1,i,j,j+1} 
\over
\vev{ \al,\bet, i-1,i }_i \vev{ \al,\bet, j-1,j }_j \vev{ \al,\bet, j, j+1 }_{j+1}
}
\nn\\
&&\hskip 3mm
+{ 
\vev{i-1,i} \vev{i,i+1,j,j+1} -\vev{i,i+1} \vev{i-1,i,j,j+1} 
\over
\vev{ \al,\bet, i-1,i }_i \vev{ \al,\bet, i, i+1 }_{i+1} \vev{ \al,\bet, j, j+1 }_{j+1}
}
\\
&&\hskip 3mm
+{ 
\vev{i,i+1} \vev{i-1,i,j-1,j}- \vev{i-1,i} \vev{i,i+1,j-1,j} 
\over
\vev{ \al,\bet, i-1,i }_i \vev{ \al,\bet, i, i+1 }_{i+1} \vev{ \al,\bet, j-1,j }_j 
}
\Bigg]
\nn\\
&=& \int \frac{d^{4}x_{0}}{i \pi^2} 
\Bigg[
{ 
{x}_{i+1,j+1}^2 - {x}_{i+1,j}^2   
\over
{\hx_{0,i+1}^2 \hx_{0,j}^2 \hx_{0,j+1}^2 }
}
{+}{ 
{x}_{i,j}^2 - {x}_{i,j+1}^2   
\over
{\hx_{0,i}^2 \hx_{0,j}^2 \hx_{0,j+1}^2 }
}
{+}{ 
{x}_{i+1,j+1}^2 - {x}_{i,j+1}^2   
\over
{\hx_{0,i}^2 \hx_{0,i+1}^2 \hx_{0,j+1}^2 }
}
{+}{ 
{x}_{i,j}^2 - {x}_{i+1,j}^2   
\over
{\hx_{0,i}^2 \hx_{0,i+1}^2 \hx_{0,j}^2 }
}
\Bigg]
\nn
\ea
and
\ba
\label{magic}
\twome  \big|_{\rm box}
&=&
\int \dZ 
{ 
\vev{i,j} \vev{\al,\bet, \bi,\bj} + \vev{\bi,\bj} \vev{\al,\bet, i,j}
\over
\vev{\al,\bet} 
\vev{ \al,\bet, i-1,i }\vev{ \al,\bet, i, i+1 }
\vev{ \al,\bet, j-1,j } \vev{ \al,\bet, j, j+1 }
} 
\nn\\
&=&
{ \vev{i,j} \vev{\bi,\bj} 
\over
\vev{i-1,i} \vev{i,i+1} \vev{j-1,j} \vev{j,j+1} }
\int \frac{d^{4}x_{0}}{i \pi^2} 
{ x_{0a}^2 + x_{0b}^2 
\over
x_{0i}^2 x_{0,i+1}^2 x_{0j}^2  x_{0,j+1}^2 } 
\ea
where $x_a$ and $x_b$
are the two solutions of the equations
$x_{0i}^2 = x_{0,i+1}^2 = x_{0j}^2  = x_{0,j+1}^2 =0$,
and where in \eqn{magic} we have 
used the identities 
\be
N_{ij} 
=
\vev{i-1, i, i+1, j} \vev{i, j-1, j, j+1}
=\vev{i,j,\bi,\bj}   \, .
\ee
The presence of two-brackets in 
$\twome  \big|_{\rm tri}$ and $\twome  \big|_{\rm box}$
indicate that these expressions are not individually dual conformal invariant.
This is not surprising, as scalar triangle diagrams violate dual
conformal invariance.

We observe that the integrands in  \eqn{magic}
contain ``magic'' numerators,
which render the resulting integrals IR-finite. 
Hence we have dropped the mass dependence in the denominator.
One can show that this integral is in fact equivalent to
the scalar two-mass easy integral in six dimensions \cite{Dixon:2011ng}
\be
\twome  \big|_{\rm box}
= 
\twomesix 
\equiv
\frac{2}{x_{ab}^2}  
\int  {d^6 x_0   \over i \pi^3}
{
 x_{ij}^2 x_{i+1,j+1}^2 -x_{i+1,j}^2 x_{i,j+1}^2 
\over
x_{0i}^2 x_{0,i+1}^2 x_{0j}^2  x_{0,j+1}^2 }\,.
\ee
Since this integral is IR-finite, it is manifestly
independent of which IR regulator is used to regulate
the amplitude.
In particular, it has no dependence on $\alpha_i$ in 
the uniform small mass limit of 
the differential-mass Higgs regulator
introduced in sec.~\ref{sect-diff}.

\vfil\break

\end{document}